\documentclass[12pt]{article}
\usepackage[lmargin=2.0cm,rmargin=2.0cm,tmargin=2.0cm,bmargin=2.5cm]{geometry}
\usepackage[T1]{fontenc}
\usepackage[latin1]{inputenc}
\usepackage[english]{babel}
\usepackage{moreverb}
\usepackage[pdftex]{graphicx}
\usepackage[hang]{subfigure}
\usepackage{amsfonts, amssymb}
\usepackage{amsmath}
\usepackage{amsthm}
\usepackage{a4}
\usepackage{float}
\usepackage{epsfig}
\usepackage{array}
\usepackage{url}
\usepackage{natbib}
\usepackage{mathabx}
\usepackage[usenames,dvipsnames]{color}
\usepackage{algorithmic}

\definecolor{dunkelmagenta}{rgb}{.3, 0, .3}

\bibpunct{(}{)}{;}{a}{,}{,}

\theoremstyle{remark}

\newtheorem{definition}{Definition}

\usepackage[raggedright]{titlesec}

\usepackage{graphicx}
\graphicspath{ {./images/sa/} }
\usepackage[export]{adjustbox}
\usepackage{caption}
\usepackage{subcaption}

\usepackage{rotating}

\usepackage{tikz}
\usepackage{tkz-graph}  
\usetikzlibrary{shapes.geometric}
\usetikzlibrary{bayesnet, positioning}
\tikzset{ 
    solid edge/.style={->, thick, color=black},
    dashed edge/.style={->, dashed, color=black},
}

\usepackage{bm}
\usepackage{amsmath}
\usepackage{mathtools}
\usepackage{amssymb}
\usepackage{amsthm}
\usepackage{nicematrix}

\usepackage[section]{placeins}

\usepackage{xcolor}
\usepackage{tcolorbox}
\usepackage[inline]{enumitem}
\usepackage{soul}

\usepackage{algorithm}
\usepackage{algorithmic}
\usepackage{float}
\newfloat{algorithm}{t}{lop}

\newtheorem{proposition}{Proposition}

\begin{document}

\begin{center}
{\large \bf An MCMC hypothesis test to check a claimed sampler: applied to a claimed sampler for the G-Wishart distribution}

\large \bf H\aa kon Tjelmeland and Hanna Bu Kval\o y\\
\large \bf Department of Mathematical Sciences\\ Norwegian University of Science and Technology\\
\large \bf Trondheim, Norway
\end{center}

\begin{abstract}
  Suppose we have a distribution of interest, with density $p(x),x\in {\cal X}$ say,  and an algorithm
  claimed to generate samples from $p(x)$. Moreover, assume we have available a Metropolis--Hastings
  transition kernel fulfilling detail balance with respect to $p(x)$. In such a situation we
  formulate a hypothesis test where $H_0$ is that the claimed sampler really generates
  correct samples from $p(x)$. We use that if initialising the Metropolis--Hastings algorithm with a
  sample generated by the claimed sampler and run the chain for a fixed number of updates, the
  initial and final states are exchangeable if $H_0$ is true. Combining this idea with the
  permutation strategy we define a natural test statistic and a valid p-value.

  Our motivation for considering the hypothesis test situation is a proposed sampler in the
  literature, claimed to generate samples from G-Wishart distribution. As no proper
  proof for the validity of this sampler seems to be available, we are exactly in the hypothesis
  test situation discussed above. We therefore apply the defined hypothesis test to the
  claimed sampler. For comparison we also apply the hypothesis test to a known exact sampler
  for a subset of G-Wishart distributions. The obtained p-values clearly show that
  the sampler claimed to be able to generate samples from any G-Wishart distribution is
  in fact not sampling from the specified distribution. In contrast, and as one should expect,
  the p-values obtained when using the known exact algorithm does not indicate any problems.
\end{abstract}

%


\section{Introduction}

Suppose we have a distribution of interest, with density $p(x),x\in{\cal X}$ say, an algorithm
claimed to generate samples from $p(x)$, and an associated computer code implementation
of the given algorithm. It is then natural to try to verify empirically that the implementation
of the algorithm is correctly generating samples from the specified distribution
$p(x)$. Such a situation can be formalised as a statistical hypothesis test, where the null
hypothesis, $H_0$, is that the algorithm and the associated implementation is correctly
generating independent samples from $p(x)$, and the alternative is that the generated samples
are from from some distribution that differ from $p(x)$. A test statistic would naturally
be defined as a function of $s$ (say) samples generated by the given algorithm. What test
statistic to use would depend on our knowledge about $p(x)$. For example, if we for
some function $g(x),x\in{\cal X}$ have available a formula for the mean $E_p[g(x)]$,
a natural choice would be to let a test statistic measure the difference between
$E_p[g(x)]$ and the empirical mean of the generated samples. If an alternative
simulation algorithm is available for generating samples from $p(x)$, and we trust the
correctness of the alternative algorithm and its implementation, a natural
alternative would be to generate $s$ independent samples by each of the two algorithms
and using the permutation method to define a test statistic. Exact permutation test for
such a situation are among other places discussed in \citet{art197} and \citet{art198}.
In these articles it is assumed that the samples are observed, but the same hypothesis
test setup is clearly valid also when the samples are generated by simulation
algorithms on a computer. In this situation, such a permutation test would naturally
be considered as a Monte Carlo test \citep{art199,art200}.

In the present article our focus is on the hypothesis test situation described above
when neither any helpful analytical properties of $p(x)$ nor any alternative simulation
algorithms are available. However, we assume that a Markov chain fulfilling detailed
balance with respect to $p(x)$ is available, typically constructed by the
Metropolis--Hastings setup. As our goal is potentially to invalidate a proposed
simulation algorithm we find it natural to limit the attention to exact tests
and associated valid p-values, and we do this by adopting the permutation method.

\citet{art196} are the first to point out that the permutation strategy
can be combined with Monte Carlo testing to define exact tests also when independent
samples can not be generated, it is sufficient that the generated samples are exchangeable.
The setup introduced in \citet{art196} is detailed and generalised in \citet{tech36}. In
both of these articles, however, the focus is on a situation where one (or a few) samples
are observed, and the goal is to test whether these samples could be coming from a specified
distribution $p(x)$. Whether the samples are observed or generated by an algorithm, does not
fundamentally change the hypothesis test situation. The only difference is that when the samples
are generated by an algorithm, a large number of samples can easily be generated. Thereby
it becomes less important to get as much power as possible out of each sample.
High power can in stead be obtained by increasing the number of samples.

Our main motivation for studying the above defined hypothesis test situation, is to check
a simulation algorithm introduced in \citet{art188}. The algorithm is claimed to generate
correct samples from the so called G-Wishart distribution, which is the standard choice of prior
for the precision matrix in a Gaussian graphical model \citep{art184,art183,art182,book39}.
The G-Wishart distribution is parametrised by a graph and in a Bayesian Gaussian graphical
model a prior is assumed for the graph. A G-Wishart prior is adopted for the precision
matrix given the graph, and a Gaussian likelihood is assumed for the observed data. A
complication when considering the resulting posterior distribution is that the normalising
constant for the G-Wishart distribution, which is a function of the graph, is in general
computationally intractable. The posterior distribution is thereby doubly intractable
\citep{art120,phd5,pro20}, so the basic Metropolis--Hastings setup can not be used
to simulate from the posterior. The standard solution for sampling from a doubly
intractable posterior distribution is to adopt the exchange
algorithm of \citet{pro20}, but this require that a sampler for the G-Wishart distribution
is available. The introduction of the sampler of \citet{art188} was therefore very important,
as for a general graph no other sampling algorithm for the G-Wishart distribution is
presently known.
\citet{art183}, \citet{art194} and \citet{art184} propose variants of the exchange algorithm
for the posterior distribution for Gaussian graphical models, all based on the
presence of the sampler of \citet{art188}. However, \citet{art188} provide no proper proof
of the correctness of the proposed sampler. The article provides some arguments for the
validity of the sampler, but it is not clear how these arguments can be expanded to a detailed
proof. We are therefore in the hypothesis test situation discussed above.

The remaining part of the present article is organised as follows. In Section \ref{sec:MH}
we discuss how an exact sampler can be combined with a Metropolis--Hastings setup to
generate exchangeable variables. Thereafter, in Section \ref{sec:test}, we use this to
define our hypothesis test setup. We introduce the G-Wishart distribution and discuss its
properties in Section \ref{sec:G-Wishart}. We start this section by a review of some
graph concepts that we need in our discussion of the G-Wishart distribution, then define
the G-Wishart distribution, discuss a block Gibbs sampling algorithm for it, formulate
an exact sampler for G-Wishart distributions defined with respect to a decomposable graph
\citep{book34}, and finally describe the proposed sampler of \citet{art188}.
In Section \ref{sec:numerical results} we provide numerical results when applying the
proposed hypothesis test on the formulated exact sampler for decomposable graphs and
on the proposed sampler of \citet{art188}. Finally, in Section
\ref{sec:closing} we provide some closing remarks.


\section{\label{sec:MH}Metropolis--Hastings algorithm and exchangeable variables}
In this section we describe how an exact sampler can be combined with the Metropolis--Hastings setup
to generate two exchangeable samples from a target distribution. Our
setup is a simplified version of what is described in \citet{art196} and \citet{tech36}.

Let $x\in{\cal X}$ be a random vector with some specified distribution $p(dx)$. The ingredients of the
basic Metropolis-Hastings scheme for simulating from $p(dx)$ is a proposal kernel $R(dy|x)$ and an
acceptance probability
\begin{align}\label{eq:alpha}
  \alpha(y|x) = \min\left\{1, \frac{p(dy)}{p(dx)} \cdot \frac{R(dx|y)}{R(dy|x)}\right\}.
\end{align}
The Metropolis--Hastings algorithm simulates a Markov chain. Letting $x$ denote the current state of the
Markov chain, the simulation of one step of the Markov chain consists of first sampling a potential new
state $y$ from $R(dy|x)$, and thereafter accepting $y$ as the new state with probability $\alpha(y|x)$.
If the proposal is not accepted, the state of the Markov chain remains unchanged. The
transition kernel of the resulting Markov chain is 
\begin{align}\label{eq:P}
P(dy|x) = R(dy|x)\alpha(y|x) + \delta_x(dy)\int_{u\in \cal X} (1-\alpha(u|x))R(du|x),
\end{align}
where $\delta_x(dy)$ equals one if $x\in dy$, and is zero otherwise. It is easy to show that this
transition kernel fulfils detailed balance with respect to $p(dx)$,
\begin{align}\label{eq:dbc}
p(dx)P(dy|x) = p(dy)P(dx|y),
\end{align}
see for example the discussion in \citet{art201}. From detailed balance it in turn follows that 
\begin{align}\label{eq:stationary}
\int_{y\in\cal X} p(dy)P(dx|y) = p(dx),
\end{align}
which means that $p(dx)$ is a stationary distribution of the simulated Markov chain. If the proposal
kernel $R(dy|x)$ is chosen so that the Markov chain is irreducible, aperiodic and recurrent, $p(dx)$ is
also the unique limiting distribution of the Markov chain, see for example the discussion in \citet{book44}.

Detailed balance is closely related to the concept of exchangeable random variables.
\begin{definition}[Exchangeable variables]
  Two random variables $x$ and $y$ are said to be exchangeable if the joint distribution of $(x,y)$ is
  identical to the joint distribution of $(y,x)$.
\end{definition}
\begin{proposition}\label{prop:exchangeable}
  Let $P(dy|x)$ be a transition kernel fulfilling detailed balance
  with respect to $p(dx)$, and let $x\sim p(dx)$ and $y|x\sim P(dy|x)$. Then $x$ and $y$ are exchangeable.
\end{proposition}
The proposition follows directly from (\ref{eq:dbc}). The result in this proposition means that if we
initialise a Metropolis--Hastings algorithm with a sample $x$ from $p(dx)$, and let $y$ be the
resulting state after one iteration, then $x$ and $y$ are exchangeable. In fact, this result is valid
also when $y$ is the state after more than one iteration. To see this, first define
the $r$-update transition kernel,
\begin{align}
  P^r(dy|x) = \idotsint P(du_1|x) \left[ \prod_{i=2}^{r-1} P(du_i|u_{i-1})\right]P(dy|du_{r-1}),
\end{align}
where the integration is over $(u_1,\ldots,u_{r-1})\in{\cal X}^{n-1}$. Assuming $P(dy|x)$ to fulfil
detailed balance with respect to $p(dx)$ we get
\begin{align}
  p(dx)P^r(dy|x) &= \idotsint p(dx) P(du_1|x) \left[ \prod_{i=2}^{r-1} P(du_i|u_{i-1})\right]P(dy|du_{r-1})\\
  &= \idotsint p(dy) P(du_{r-1}|y) \left[\prod_{i=2}^{r-1} P(du_{i-1}|du_i)\right] P(dx|u_1)\\
  &= p(dy) P^r(dx|y),
\end{align}
where we first move $p(dx)$ inside the integrals, then use (\ref{eq:dbc}) repeatedly, and finally
move $p(dy)$ outside the integrals and recognise the integrals as being identical to $P^r(dx|y)$. So
if $P(dy|x)$ fulfils detailed balance with respect to $p(dx)$, so do $P^r(dy|x)$. Thereby, as one step
with $P^r(dy|x)$ is equivalent to $r$ Metropolis--Hastings updates with $P(dy|x)$, Proposition \ref{prop:exchangeable}
gives the result.

In the above discussion we have assumed a Metropolis--Hastings algorithm based on only one
proposal kernel $R(dy|x)$. Most Metropolis--Hastings algorithms used in practice are based on
a combination of several proposal kernels; $R_k(dy,x),k=1,\ldots,K$ say. For each $R_k(dy|x)$
a corresponding acceptance probability $\alpha_k(y|x)$ is defined according to (\ref{eq:alpha}).
Proposing a potential new state according to $R_k(dy|x)$ and accepting it with probability
$\alpha_k(y|x)$ corresponds to using a transition kernel $P_k(dy|x)$ defined as in (\ref{eq:P}).
As explained above, each $P_k(dy|x)$ is by
construction fulfilling detailed balance with respect to $p(dx)$. Moreover, for any $k=1,\ldots,K$, a Markov
chain with transition kernel $P_k(dy|x)$ have $p(dx)$ as a stationary distribution. So replacing $P(dy|x)$
with $P_k(dy|x)$ in (\ref{eq:dbc}) and (\ref{eq:stationary}), the two expressions are still correct.

There are different ways to combine $P_k(dy|x),k=1,\ldots,K$
to define a Markov chain. The perhaps most popular strategy is first to do one update according to $P_1(dy|x)$,
thereafter do one update according to $P_2(dy|x)$, and continue like this up to $P_k(dy|x)$, thereafter
starting with an update
according to $P_1(dy|x)$ again, and so on. More generally, one can choose some fixed permutation
$\sigma=(\sigma(1),\ldots,\sigma(K))$ of the integers from $1$ to $K$, and first do an
update according to $P_{\sigma(1)}(dy|x)$, then
an update according to $P_{\sigma(2)}(dy|x)$ and so on. The resulting transition kernel for the full
sweep of $P_{\sigma(k)}(dy|x),k=1,\ldots,K$ becomes
\begin{align}\label{eq:Psigma}
  P_\sigma(dy|x) = \idotsint P_{\sigma(1)}(du_1|x) \left[\prod_{k=2}^{K-1}P_{\sigma(k)}
    (du_k|u_{k-1})\right] P_K(dy|u_{K-1}),
\end{align}
where the integration is again
over $(u_1,\ldots,u_{k-1})\in{\cal X}^{k-1}$. Using that (\ref{eq:stationary}) is valid for
each $P_k(dy|x)$, it follows that (\ref{eq:stationary}) is also valid when $P(dy|x)$ is replaced by $P_\sigma(dy|x)$.
So $p(dx)$ is a stationary distribution for the Markov chain defined by $P_\sigma(dy|x)$.
Note, however, that $P_\sigma(dy|x)$ does not fulfil detailed balance with respect to $p(dx)$,
which means that it can not be used to generate exchangeable variables as discussed above. In the
following we discuss two alternative ways to combine $P_k(dy|x),k=1,\ldots,K$, 
for which the associated transition kernel do fulfil detailed balance with respect to $p(dx)$.

The conceptually and implementationally perhaps simplest way to combine the $P_k(dy|x),k=1,\ldots,K$ is 
in each step of the Markov chain to draw uniformly at random, independently at each step, which transition kernel $P_k(dy|x)$
to use. The resulting transition kernel of one update using this random update strategy, $P_{\mbox{\tiny ru}}(dy|x)$, is
\begin{align}\label{eq:ru}
P_{\mbox{\tiny ru}}(dy|x) = \frac{1}{K}\sum_{k=1}^K P_k(dy|x).
\end{align}
It is straight forward to show that $P_{\mbox{\tiny ru}}(dy|x)$ indeed fulfils detailed balance with respect to $p(dx)$. 

An alternative construction of a transition kernel which fulfils detailed balance with respect to $p(dx)$ is the following.
At each iteration of the Markov chain, draw uniformly at random, independently in each iteration, a permutation $\sigma$ of the
integers $1$ to $K$ and do one update according to $P_\sigma(dy|x)$ defined in (\ref{eq:Psigma}). The resulting
transition kernel for one iteration of such a random permutation strategy, $P_{\mbox{\tiny rp}}(dy|x)$, is
\begin{align}\label{eq:rp}
P_{\mbox{\tiny rp}}(dy|x) = \frac{1}{K!} \sum_{\sigma}P_\sigma(dy|x),
\end{align}
where the sum is over all possible permutations. Again it is straight forward to show that $P_{\mbox{\tiny rp}}(dy|x)$
fulfils detailed balance with respect to $p(dx)$.

The essential conclusion from the above discussion is the following. If we initialise a Metropolis--Hastings algorithm
with an $x$ sampled from $p(dx)$, run $r$ iterations according to $P_{\mbox{\tiny ru}}(dy|x)$ or $P_{\mbox{\tiny rp}}(dy|x)$,
and let $y$ denote the final state, then $x$ and $y$ are exchangeable.

\section{\label{sec:test}MCMC hypothesis test setup}

Using ideas introduced in \citet{art196}, and also discussed in \citet{tech36}, we here formulate
a Monte Carlo hypothesis test setup that can be used to test whether a claimed exact sampler
is really producing samples from the specified distribution.

Assume we have a random variable $x\in {\cal X}$ of interest with distribution $p(dx)$, and an associated
proposed simulation algorithm that is claimed to produce exact samples from $p(dx)$. Moreover, assume we
have available a Metropolis--Hastings algorithm with a transition kernel $P(dy|x)$ that fulfils
detailed balance with respect to $p(dx)$, as discussed in Section \ref{sec:MH}.
So, as discussed in Section \ref{sec:MH}, if we generate an exact sample $x^{(0)}$ from $p(dx)$ and
use this as the initial state in the Metropolis--Hastings algorithm and run $r$ (say) steps with
the chosen $P(dy|x)$ to generate $x^{(r)}$, we have that $x^{(0)}$ and $x^{(r)}$ are exchangeable.
We now want to use such a simulation setup to test the null hypothesis
\begin{align}
  H_0: \mbox{The proposed algorithm is producing exact samples from $p(dx)$}.
\end{align}
To obtain a test statistic that can be used to test this null hypothesis we use the following scheme. First
we use the proposed simulation algorithm to generate $s$ independent samples $x_i^{(0)},i=1,\ldots,s$.
Under $H_0$ these samples are exact samples from $p(dx)$. Second, for each $i=1,\ldots,s$
we initiate the Metropolis--Hastings algorithm with $x_i^{(0)}$ and run $r$ steps with $P(dy|x)$ to obtain $x_i^{(r)}$.
Under $H_0$, also each $x_i^{(r)}$ is an exact sample from $p(dx)$, and $x_i^{(0)}$ and $x_i^{(r)}$ are
exchangeable. If $H_0$ is not true, the Metropolis--Hastings algorithm have a burn-in period, so the distribution
of $x_i^{(r)}$ differ from the distribution of $x_i^{(0)}$. 
Third, for some chosen function $h: {\cal X}\rightarrow\mathbb{R}$ we define $t_{i1}=h(x_i^{(0)})$
and $t_{i2}=h(x_i^{(r)})$ for $i=1,\ldots,s$, and organise these values in an $s\times 2$ matrix $T=[t_{ij}]$.
The rows of $T$ are by construction independent, and under $H_0$ the two elements in the same row,
$t_{i1}$ and $t_{i2}$, are by construction exchangeable. Thus, under $H_0$ the joint distribution of the
elements of $T$ remains unchanged if interchanging $t_{i1}$ and $t_{i2}$ in one or more rows $i$.
Last, we define the test statistic to be
\begin{align}
  \omega^\star =  H(T),
\end{align}
where $H$ is some scalar function of $T$. By construction all elements in the first column of $T$
are independent samples from the same distribution, and likewise all elements in the second column
are independent samples from the same distribution. If $H_0$ is true the distributions for the
two columns are identical, whereas if $H_0$ is not true the two distributions are different. In the following
we assume the function $H$ is chosen so that a difference in the distributions for the two
columns of $T$ tends to give high values for $H(T)$. Thus, observing
a high value for $\omega^\star$ indicates that $H_0$ is not true.

To obtain a p-value associated to the observed value of $\omega^\star$ we adopt a permutation strategy, see
for example the discussion in \citet{art197}. We generate $q$ (say) variants of the test statistic by
randomly permuting the two elements in the same row of $T$. More specifically, we first draw
$v_{ij}\in\{0,1\},i=1,\ldots,s,k=1,\ldots,q$ independently with probability a half for each of $0$ and $1$.
Then we define
\begin{align}
  t_{i1}^{(k)} = v_{ik} t_{i1} + (1-v_{ik}) t_{i2} \mbox{~~~and~~~} t_{i2}^{(k)} = (1-v_{ik})t_{i1} + v_{ik}t_{i2}
\end{align}
for $i=1,\ldots,s$ and $k=1,\ldots,q$, and the $k$'th re-sampled variant of $T$ is $T^{(k)}=[t_{ij}^{(k)}]$.
The associated value of the test statistic is
\begin{align}
  \omega^{(k)} = H(T^{(k)}).
\end{align}
From the above discussion it follows that under $H_0$ the random variables $\omega^\star,\omega^{(1)}$, $\ldots,\omega^{(q)}$
are exchangeable. Thus, the probability that $\omega^\star$ is in the top $\alpha$ proportion of
$\omega^\star,\omega^{(1)},\ldots,\omega^{(q)}$ is, by symmetry, at most $\alpha$. So if we define
\begin{align}
  p^\star = \frac{1}{q+1}\left( 1 + \sum_{k=1}^q \mathbb{I}(\omega^{(k)} \geq \omega^\star)\right),
\end{align}
where $\mathbb{I}(A)$ is the indicator function being equal to one if $A$ is true and zero
otherwise, we have
\begin{align}
  P\left(\left. p^\star \leq \alpha \right| \mbox{$H_0$ is true}\right) \leq \alpha.
\end{align}
So $p^\star$ is a valid p-value.

\section{\label{sec:G-Wishart}The G-Wishart distribution}
In this section we start by introducing some graph concepts, which we thereafter use to define and discuss
properties of the G-Wishart distribution. In particular we formulate an exact sampler for the G-Wishart
distribution when defined with respect to a decomposable graph, and the claimed general sampler
of \citet{art188}.

\subsection{\label{sec:graph}Some graph concepts}
Here we define what we mean by an undirected graph, and define and discuss some related concepts that
we need in our discussion of the G-Wishart distribution. A detailed
introduction to graph theory can for example be found in \citet{book34} and \citet{book39}.

\begin{figure}
  \begin{center}
    \begin{tabular}{ccccccc}
      \begin{tikzpicture}[scale=0.25]
        \draw[thick](0,0) circle(1cm) node{$3$};
        \draw[thick](4,0) circle(1cm) node{$4$};
        \draw[thick](0,4) circle(1cm) node{$1$};
        \draw[thick](4,4) circle(1cm) node{$2$};
        
        \draw[thick] (1,0) -- (3,0);
        \draw[thick] (0,1) -- (0,3);
        \draw[thick] (1,4) -- (3,4);
        \draw[thick] (4,1) -- (4,3);
        \draw[thick] (3.293,3.293) -- (0.707,0.707);
      \end{tikzpicture}
      & \hspace*{0.25cm} &
      \begin{tikzpicture}[scale=0.25]
        \draw[thick](0,0) circle(1cm) node{$3$};
        \draw[thick](4,0) circle(1cm) node{$4$};
        \draw[thick](0,4) circle(1cm) node{$1$};
        \draw[thick](4,4) circle(1cm) node{$2$};
        
        \draw[thick] (1,0) -- (3,0);
        \draw[thick] (0,1) -- (0,3);
        \draw[thick] (1,4) -- (3,4);
        \draw[thick] (4,1) -- (4,3);
      \end{tikzpicture}
      & \hspace*{0.25cm} &
      \begin{tikzpicture}[scale=0.25]
        \draw[thick](0,0) circle(1cm) node {$8$};
        \draw[thick](4,0) circle(1cm) node {$9$};
        \draw[thick](8,0) circle(1cm) node {$10$};

        \draw[thick](0,4) circle(1cm) node {$4$};
        \draw[thick](4,4) circle(1cm) node {$5$};
        \draw[thick](8,4) circle(1cm) node {$6$};
        \draw[thick](12,4) circle(1cm) node {$7$};
        
        \draw[thick](0,8) circle(1cm) node {$1$};
        \draw[thick](4,8) circle(1cm) node {$2$};
        \draw[thick](8,8) circle(1cm) node {$3$};

        \draw[thick](1,8) -- (3,8);
        \draw[thick](0,7) -- (0,5);
        \draw[thick](0.707,4.707) -- (3.293,7.293);

        \draw[thick](0,3) -- (0,1);
        \draw[thick](1,0) -- (3,0);
        \draw[thick](0.707,3.293) -- (3.293,0.707);

        \draw[thick](5,8) -- (7,8);
        \draw[thick](5,4) -- (7,4);
        \draw[thick](4,7) -- (4,5);
        \draw[thick](8,7) -- (8,5);
        \draw[thick](4.707,4.707) -- (7.293,7.293);
        \draw[thick](4.707,7.293) -- (7.293,4.707);

        \draw[thick](8,3) -- (8,1);
        \draw[thick](9,4) -- (11,4);
        \draw[thick](8.707,0.707) -- (11.293,3.293);
      \end{tikzpicture}
      & \hspace*{0.25cm} &
      \begin{tikzpicture}[scale=0.25]
        \draw[thick](0,0) circle(1cm) node {$8$};
        \draw[thick](4,0) circle(1cm) node {$9$};
        \draw[thick](8,0) circle(1cm) node {$10$};

        \draw[thick](0,4) circle(1cm) node {$4$};
        \draw[thick](4,4) circle(1cm) node {$5$};
        \draw[thick](8,4) circle(1cm) node {$6$};
        \draw[thick](12,4) circle(1cm) node {$7$};
        
        \draw[thick](0,8) circle(1cm) node {$1$};
        \draw[thick](4,8) circle(1cm) node {$2$};
        \draw[thick](8,8) circle(1cm) node {$3$};

        \draw[thick](1,0) -- (3,0);
        \draw[thick](5,0) -- (7,0);
        \draw[thick](5,4) -- (7,4);
        \draw[thick](9,4) -- (11,4);
        \draw[thick](1,8) -- (3,8);
        \draw[thick](5,8) -- (7,8);

        \draw[thick](0,1) -- (0,3);
        \draw[thick](0,5) -- (0,7);
        \draw[thick](4,1) -- (4,3);
        \draw[thick](8,1) -- (8,3);
        \draw[thick](8,5) -- (8,7);

        \draw[thick](0.707,4.707) -- (3.293,7.293);
        \draw[thick](4.707,4.707) -- (7.293,7.293);
        \draw[thick](8.707,0.707) -- (11.293,3.293);
        \draw[thick](8.707,7.293) -- (11.293,4.707);
      \end{tikzpicture}
      \\[0.2cm]
      (a) & & (b) & & (c) & & (d)
    \end{tabular}
  \end{center}
  \caption{\label{fig:graph}Four example graphs: (a) A small decomposable graph with four nodes and five edges, (b)
    a small non-decomposable graph with four nodes and four edges, (c) a decomposable graph with ten nodes and
    fifteen edges, and (d) a non-decomposable graph with ten nodes and fifteen edges.}
\end{figure}
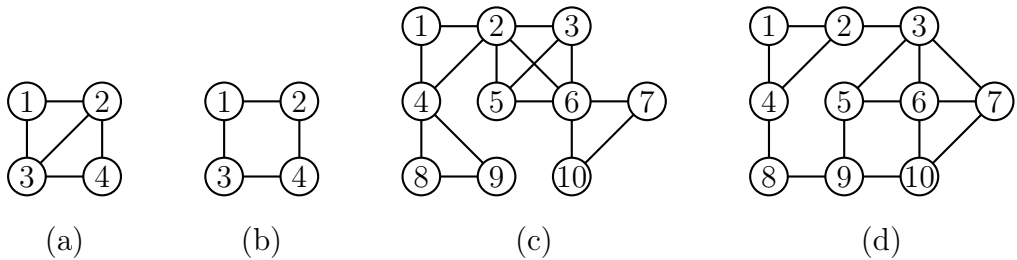

Let $G=(V,E)$ be an undirected graph, where $V$ is a finite set of nodes,
and $E\subseteq \{(i,j):i,j\in V,i < j\}$ is an associated set of edges.
Four example graphs are visualised in Figure \ref{fig:graph}.
Given such an undirected graph $G=(V,E)$ with $V=\{1,\ldots,p\}$,
we let $\mathbb{P}_G$ denote the set of all $p\times p$ symmetric and positive definite
matrices that have element $(i,j), i<j$ equal to zero whenever
both $(i,j)\not\in E$. So for $Q\in \mathbb{P}_G$, a non-diagonal
element of $Q$, $Q_{ij}$ say, is allowed to be non-zero only if $(i,j)\in E$ or $(j,i)\in E$,
in which case we say that nodes
$i$ and $j$ are neighbours. A set of nodes $C\subseteq V$ is called a clique if all distinct pairs of nodes in $C$ are
neighbours. For the graph visualised in Figure \ref{fig:graph}(a), for example $\{1,2\}$ and $\{2,4\}$ are cliques,
and for the graph shown in Figure \ref{fig:graph}(c) the sets $\{1,2,4\}$ and $\{2,3,5,6\}$ are
cliques. One should note that the definition of a clique implies that 
we have no zero-constraints for a sub-matrix of $Q$ that is obtained by picking out only the rows and columns in
$Q$ that are associated to nodes in a clique $C$. Such a sub-matrix we denote by $Q_{C,C} = [Q_{ij}; i,j\in C]$. If a clique
is not a proper subset of another clique we say that it is a maximal clique. The graph visualised in Figure
\ref{fig:graph}(a) has two maximal cliques, namely $\{1,2,3\}$ and $\{2,3,4\}$, whereas
the graph shown in Figure \ref{fig:graph}(c) has the four maximal cliques
$\{1,2,4\}, \{2,3,5,6\}, \{4,8,9\}$ and $\{6,7,10\}$.

For an undirected graph $G=(V,E)$ and a subset of nodes $A\subseteq V$, we say that the graph $G_A=(A,E_A)$,
where $E_A=\{(i,j)\in E:i,j\in A\}$, is the subgraph induced by $A$. For the graph in Figure \ref{fig:graph}(c)
we have for example that the subgraph induced by $A=\{1,2,4,5\}$ is $G_A = (A,\{(1,2),(1,4),(2,4),(2,5)\})$.
For some undirected graphs there exists one or more proper decompositions, which is defined as follows.
\begin{definition}[Proper decomposition]
  Let $G=(V,E)$ be an undirected graph. Then a triple $(A,B,C)$ of non-empty and disjoint subsets of $V$ is said to
  form a proper decomposition of $G$ if $i$) $A\cup B\cup C=V$, $ii$) $C$ is a clique, and $iii$) none of the nodes in $A$
  is a neighbour of any of the nodes in $B$, i.e. $E\cap \{(i,j):i\in A, j\in B \mbox{ or }i\in B, j\in A\}=\emptyset$.
\end{definition}
For the graph in Figure \ref{fig:graph}(c), for example $A=\{1,4,8,9\}, B=\{3,5,6,7,10\}$ and $C=\{2\}$ form a proper
decomposition of $G$. The triple $A=\{1,4,8,9\}, B=\{7,10\}$ and $C=\{2,3,5,6\}$ forms another proper
decomposition of the same graph. An important subset of undirected graphs are called decomposable.
\begin{definition}[Decomposable graph]
  An undirected graph $G=(V,E)$ is said to be decomposable if either $i$) $V$ is a clique or $ii$) there exists a
  triple $(A,B,C)$ that forms a proper decomposition of $G$ so that the induced subgraphs $G_A$ and $G_B$ are both
  decomposable. 
\end{definition}
Of the four graphs shown in Figure \ref{fig:graph}, the graphs in (a) and (c) are decomposable, whereas the graphs in (b) and (d)
are not. The concept of decomposable graphs is important as many operations can efficiently be done for decomposable
graphs, but not for undirected graphs in general. In Section \ref{sec:decomposable} we in particular formulate an exact sampler for a
G-Wishart distribution defined with respect to a decomposable graph. To develop such an exact sampler we need a last graph concept,
that is closely related to decomposable graphs.
\begin{definition}[Perfect ordering of nodes]
  Let $G=(V,E)$ be an undirected graph with $V=\{1,\ldots,p\}$. Then a numbering of the nodes, $(v_1,\ldots,v_p)$ say, is said to be perfect
  if for each node $i\in V$ the set of the neighbours of node $i$ that has a lower number than $v_i$ is a clique.
\end{definition}
For the graphs in Figures \ref{fig:graph}(a) and (c), $v_i=i,i=1,\ldots,p$ is a perfect ordering of the nodes, whereas for the graphs in (b)
and (d) there exist in fact no perfect ordering of the nodes. As discussed in \citet{book34} and proved in \citet{book39}, the concepts
of a decomposable graph and a perfect ordering of the nodes are closely related in that an undirected graph $G$ is decomposable if and
only if there exists a perfect ordering of the nodes. Algorithms for deciding whether a given undirected graph is decomposable and,
if it is decomposable, find a perfect ordering of the nodes are given in \citet{book34}.

\subsection{\label{sec:GW}The G-Wishart distribution and block Gibbs sampling}

For an undirected graph $G=(V,E)$ with $V=\{1,\ldots,p\}$, the G-Wishart distribution \citep{art191,art189} defines a
distribution for $Q\in \mathbb{P}_G$ with density
\begin{align}\label{eq:GW}
  f(Q|\delta,D,G) = c(\delta,D,G) |Q|^{\frac{\delta}{2}-1}\exp\left\{ -\frac{1}{2}\mbox{tr}(QD)\right\} \mathbb{I}(Q\in \mathbb{P}_G),
\end{align}
where $\delta>0$ is a scalar parameter, $D\in \mathbb{R}^{p\times p}$ is a positive definite matrix parameter, $\mbox{tr}(QD)$
is the trace of $QD$, and $c(\delta,D,G)$ is the normalising constant. One should note that with
the requirements that $Q$ should
be symmetric and have $Q_{ij}=0$ for all $(i,j)\not\in E, i<j$, the number of free parameters in $Q$ is
$p+|E|$, where $|E|$ denotes the number of edges in $G$. The formula in (\ref{eq:GW}) should be understood as a density
for these free parameters. When a matrix $Q$ is distributed according to (\ref{eq:GW}) we write
$Q\sim {\cal W}_G(\delta,D)$

The G-Wishart distribution is a generalisation of the Wishart distribution in that if $G$ is a complete graph, i.e.
$E=\{(i,j):i,j\in V,i<j\}$, so that we have no zero constraints on the elements of $Q$, the G-Wishart
distribution becomes a Wishart distribution. When a matrix $Q$ is distributed according to (\ref{eq:GW}) for
a complete graph $G$ we write $Q\sim {\cal W}(\delta,D)$. 

When $C$ is a clique in $G$, the conditional distribution for the sub-matrix $Q_{C,C}$
given all other elements in $Q$ is analytically available and easy to sample from.
Defining also the sub-matrices $Q_{C,V\setminus C}=[Q_{ij};i\in C,j\in V\setminus C]$ and
$Q_{V\setminus C,V\setminus C}=[Q_{ij};i,j\in V\setminus C]$, and letting
\begin{align}\label{eq:Schur}
  Q_{C,C}^\star = Q_{C,C} - Q_{C,V\setminus C}\ (Q_{V\setminus C,V\setminus C})^{-1}\ Q_{C,V\setminus C}^T
\end{align}
denote the Schur complement of $Q_{C,C}$ in $Q$, it is straightforward to show that if
$Q\sim {\cal W}_G(\delta,D)$ we have 
$Q_{C,C}^\star\sim {\cal W}(\delta,D_{C,C})$, where $D_{C,C}=[D_{ij};i,j\in C]$,
and that $Q_{C,C}^\star$ is independent of the
pair $(Q_{C,V\setminus C},Q_{V\setminus C,V\setminus C})$. Solving (\ref{eq:Schur}) with respect to
$Q_{C,C}$ we have
\begin{align}\label{eq:conditional}
  Q_{C,C} = Q_{C,C}^\star + Q_{C,V\setminus C}\ (Q_{V\setminus C,V\setminus C})^{-1}\ Q_{C,V\setminus C}^T,
\end{align}
which means that we may generate a realisation from the conditional distribution for $Q_{C,C}$ given
the pair $(Q_{C,V\setminus C},Q_{V\setminus C,V\setminus C})$ by first generating
$Q_{C,C}^\star\sim {\cal W}(\delta,D_{C,C})$ independent of everything else, and thereafter compute
$Q_{C,C}$ by (\ref{eq:conditional}). We end this section by describing how such a
conditional simulation scheme can be used to construct a Metropolis--Hastings transition
kernel that fulfils detail balance with respect to the G-Wishart density in (\ref{eq:GW}).

Assume the undirected graph $G$ has $m$ maximal cliques, denoted by $C_1,\ldots,C_m$.
For each maximal clique $C_k,k\in\{1,\ldots,m\}$, we can then define a Metropolis--Hastings
transition kernel with proposal kernel $R_k(\cdot|\cdot)$ defined by the conditional distribution
for $Q_{C_k,C_k}$ given $Q_{C_k,V\setminus C_k}$ and $Q_{V\setminus C_k,V\setminus C_k}$. As this
is a Gibbs proposal the associated acceptance probability $\alpha_k(\cdot|\cdot)$ is identically
equal to unity. As discussed in Section \ref{sec:MH}, the resulting transition kernel,
denoted by $P_k(\cdot|\cdot)$, then fulfils detailed balance with respect to the target
G-Wishart density in (\ref{eq:GW}). One can then use (\ref{eq:ru}) or (\ref{eq:rp}) to combine
the $m$ transition kernels $P_k(\cdot|\cdot),k=1,\ldots,m$ into one transition kernel
that fulfils detail balance with respect to the G-Wishart density.


\subsection{\label{sec:decomposable}Exact sampler for G-Wishart distribution when the graph is decomposable}
In this section we introduce an exact sampler for the G-Wishart distribution when the associated graph is
decomposable. Our motivation for deriving an exact sampler in this case is just to demonstrate that
the hypothesis test defined in Section \ref{sec:test} does not reject $H_0$ for this sampler.
Many theoretical properties are available for G-Wishart distributions defined on a
decomposable graph, and in particular exact and simple to compute formulas are available for the normalising
constant, so our exact sampler is not really of much value except for what we are using it for here.

As discussed in Section \ref{sec:graph}, for a decomposable graph $G=(V,E)$ with $V=\{1,\ldots,p\}$
there exists a perfect ordering of the nodes, $(v_1,\ldots,v_p)$. Without loss of generality
we assume in this section $v_i=i,i=1,\ldots,p$ to be a perfect ordering of the nodes. If this is not a
perfect ordering, one can always rename the nodes so that this becomes the case One should note
that this just corresponds to a permutation of the elements in the associated precision matrix $Q$.

Let $Q\sim {\cal W}_G(\delta,D)$ and split $Q\in \mathbb{R}^{p\times p}$
into the four blocks $Q_{11} = [Q_{ij};i,j=1,\ldots,p-1]\in \mathbb{R}^{(p-1)\times (p-1)}$,
$q_{12}=[Q_{ip};i=1,\ldots,p-1]\in\mathbb{R}^{(p-1)}$, $q_{21}=q_{12}^T$, and
$q_{22} = Q_{pp}\in \mathbb{R}$, and denote the Schur complement of $Q_{11}$ by $Q_{11}^\star$, i.e.
\begin{align}\label{eq:q11}
  Q_{11}^\star = Q_{11} - \frac{q_{12}\, q_{12}^T}{q_{22}}.
\end{align}
Make a corresponding decomposition of the matrix parameter $D$, i.e.
$D_{11}=[D_{ij};i,j = 1,\ldots,p-1]$, $d_{12}=[D_{ip};i=1,\ldots,p-1]$, $d_{21}=d_{12}^T$ and
$d_{22}=D_{pp}$. Using well known properties for the determinant and trace of positive definite matrices
one gets that the joint distribution of $Q_{11}^\star$, $q_{12}$ and $q_{22}$ can be expressed as
\begin{align}\nonumber
  &f(Q_{11}^\star,q_{12},q_{22}) \propto |Q_{11}^\star|^{\frac{\delta}{2}-1} \exp\left\{-\frac{1}{2}\mbox{tr}(Q_{11}^\star D_{11})\right\}
  \mathbb{I}(Q_{11}^\star\in \mathbb{P}_{G_{\{1,\ldots,p-1\}}})\\ \label{eq:Qqq}
  &\times \frac{1}{q_{22}^{\frac{\delta}{2}-1}}\exp\left\{ -\frac{1}{2}\left[ \frac{q_{12}^TD_{11}q_{12}}{q_{22}} + 2q_{12}^Td_{12} + q_{22}d_{22}\right]\right\}\\
  &\times 
  \mathbb{I}(q_{22} > 0, q_{12}(i) = 0,i\not\in \partial p),
\end{align}
where $\partial p = \{i\in V: (i,p)\in E\}$ is the set of neighbours of node $p$. One should note that the expression in
(\ref{eq:Qqq}) is only correct when all neighbours of node $p$ are neighbours of each other, but this property
follows from the assumed perfect ordering of the nodes. From (\ref{eq:Qqq}) we can see that $Q_{11}^\star$ is
independent of the pair $(q_{12},q_{22})$, and that
\begin{align}\label{eq:Qstar}
  Q_{11}^\star \sim {\cal W}_{G_{\{1,\ldots,p-1\}}}(\delta,D_{11}).
\end{align}
Now let $\widetilde{q}_{12}$ denote a column vector of length $|\partial p|$ that contains the free elements of $q_{12}$, i.e.
the elements of $q_{12}$ that are allowed to differ from zero.
Mathematically this can be done by setting $\widetilde{q}_{12}=Aq_{12}$ for a $|\partial p|\times (p-1)$ matrix $A$ where
exactly one element in each row is equal to one, exactly one element in column number $j$ is equal to one if $j\in \partial p$,
and all other elements of $A$ are equal to zero. Letting $\widetilde{d}_{12}=Ad_{12}$ and $\widetilde{D}_{11}=AD_{11}A^T$ 
we can re-express the factors of (\ref{eq:Qqq}) that includes $q_{12}$ or $q_{22}$ in terms of $\widetilde{q}_{12}$ and $q_{22}$.
By completing the square for $\widetilde{q}_{12}$ in the exponent one can then
find the marginal distribution for $q_{22}$ and the conditional distribution for
$\widetilde{q}_{12}$ given $q_{22}$. The marginal density for $q_{22}$ becomes
\begin{align}\label{eq:q22}
  f(q_{22}) \propto q_{22}^{\frac{\delta + |\partial p|}{2} - 1}
  \exp\left\{-q_{22}\cdot \frac{d_{22}-\widetilde{d}_{12}^T\widetilde{D}_{11}
    \widetilde{d}_{12}}{2}\right\} \mathbb{I}(q_{22}>0),
\end{align}
which we recognise as the density of a gamma distributed variable. The conditional distribution for $\widetilde{q}_{12}$ given
$q_{22}$ becomes multivariate Gaussian,
\begin{align}\label{eq:q12}
  \widetilde{q}_{12}|q_{22} \sim N\left(-q_{22}\widetilde{D}_{11}^{-1}\widetilde{d}_{12},q_{22}\widetilde{D}_{11}^{-1}\right).
\end{align}
To sample $Q\sim {\cal W}_G(\delta,D)$ can thereby be done in three steps. First, simulate $q_{22}$ according to
(\ref{eq:q22}) and $\widetilde{q}_{12}|q_{22}$ according to (\ref{eq:q12}). Note that the simulated vector $\widetilde{q}_{12}$
defines all non-zero elements of $q_{12}$, so now we also have a realisation of $q_{12}$.
Second, simulate $Q_{11}^\star$ according to (\ref{eq:Qstar}),
and lastly, compute $Q_{11}$ according to (\ref{eq:q11}) and put $Q_{11}$, $q_{12}$ and $q_{22}$ together to form
a realisation of $Q$. Of course, to sample $Q$ according to this procedure we need to sample $Q_{11}^\star$ according to
the G-Wishart distribution specified in (\ref{eq:Qstar}), so at first sight it may seem like we have not gained anything.
However, the G-Wishart distribution in (\ref{eq:Qstar}) is defined for a graph with one node less than in $G$. So thereby
we have two possibilities. Either $p=2$, in which case $Q_{11}^\star$ is scalar and the G-Wishart distribution in
(\ref{eq:Qstar}) becomes a gamma distribution, or $p>2$ and in that case we can use the procedure just defined to
sample from the G-Wishart distribution in (\ref{eq:Qstar}).

\subsection{\label{sec:claimed}Claimed general sampler of \citet{art188}}
The claimed general sampler for a G-Wishart distribution introduced in \citet{art188} is based on
(\ref{eq:conditional}) and closely related to the Gibbs sampling algorithm mentioned above.
Here we first describe the algorithm and thereafter briefly discuss the arguments that \citet{art188}
puts forward to argue that it generates a sample according to the specified G-Wishart
distribution.

The proposed algorithm for generating a sample from (\ref{eq:GW}) starts by sampling a $\widetilde{Q}\sim
{\cal W}(\delta,D)$ and computing the corresponding inverse $\Sigma=\widetilde{Q}^{-1}$. It then uses
fix point iteration to solve with respect to $Q$ the non-linear equation system obtained by replacing
$Q_{C,C}^\star$ in (\ref{eq:conditional}) with $(\Sigma_{C,C})^{-1}$ and requiring the relation to hold for all
maximal cliques. So again letting $m$ denote the number of maximal cliques in $G$ and letting $C_1,\ldots,C_m$
denote the maximal cliques, the equation system that is solved with respect to $Q$ is
\begin{align}\label{eq:fix}
  Q_{C_k,C_k} = (\Sigma_{C_k,C_k})^{-1} + Q_{C_k,V\setminus C_k}\ (Q_{V\setminus C_k,V\setminus C_k})^{-1}\ Q_{C_k,V\setminus C_k}^T;
  ~~k=1,\ldots,m.
\end{align}
\citet{art188} claims that the resulting $Q$ is an exact sample from the G-Wishart distribution in (\ref{eq:GW}),
but does not give a proper proof for the claim. The somewhat informal arguments given in favour of the result go
in two steps. In the first step, the argument is that since $\widetilde{Q}$ is Wishart distributed,
the $\Sigma$ is inverse Wishart distributed, and by well known properties of the inverse Wishart distribution
then also $\Sigma_{C_k,C_k}$ is inverse Wishart. Thereby $(\Sigma_{C_k,C_k})^{-1}$ is Wishart distributed,
and by studying the details one will see that the parameters of this Wishart distribution is the 
same as the  parameters in the Wishart distribution for $Q_{C_k,C_k}^\star$ in (\ref{eq:conditional}).
From this \citet{art188}
concludes that for the $Q$ generated by his procedure, the conditional distribution of $Q_{C_k,C_k}$ given
all the remaining elements in $Q$ has exactly the distribution it should have according to (\ref{eq:GW}).
In the second step, \citet{art188} refers to results in \citet{art195} and \citet{art3} and concludes that
since all conditional distributions for the $Q$ generated by his procedure is equal to the corresponding
conditional distributions resulting from (\ref{eq:GW}), also the joint distributions must be equal.

We find problems in both steps of the argumentation in \citet{art188}. In the first step, it is at least not
obvious that the $Q$ generated as the solution of (\ref{eq:fix}) gives independence between a
corresponding Schur complement
$Q_{C_k,C_k}^\star$ and the pair $(Q_{C_k,V\setminus C_k},Q_{V\setminus C_k,V\setminus C_k})$ as it should for a
G-Wishart distribution. In turn, if this independence is not fulfilled, the arguments used to conclude that
the conditional distribution is as it should be is no longer valid. In the second step, even if
not explicitly writing so, \citet{art188} seems to refer to the result in the Hammersley--Clifford theorem
which is valid subject to the so-called positivity condition. The problem is
that the family of positive definite matrices does not fulfil the positivity condition.

\section{\label{sec:numerical results}Numerical results}
In our numerical experiments we consider four G-Wishart distributions, one for each of the
graphs visualised in Figure \ref{fig:graph}(a). In all cases we use $\delta=10$ and simply
set $D$ equal to the identity matrix of the correct dimension. For the G-Wishart distributions
based on the two decomposable graphs we consider both the exact simulation algorithm
derived in Section \ref{sec:decomposable} and the claimed sampler described in
Section \ref{sec:claimed}, whereas for the distributions based on the two non-decomposable graphs
we only consider the claimed sampler discussed in Section \ref{sec:claimed}.
In the following discussion we refer to the exact sampler developed in Section \ref{sec:decomposable}
as just the {\em exact sampler}, and the sampler proposed in \citet{art188} and described
in Section \ref{sec:claimed} as the
{\em claimed sampler} or {\em claimed general sampler}. 

We first describe in detail the remaining experimental setup, then
present the results of some preliminary exploration of a simulation study, and finally give the results
of formal hypothesis tests of the type introduced in Section \ref{sec:test}.

\subsection{\label{sec:setup}Experimental setup}
For each of the four G-Wishart distributions we first use the simulation algorithm to be considered
to generate $s=10\,000$ realisations $Q_i^{(0)},i=1,\ldots,s$. The claimed sampler 
is iterative and should be run until convergence. To avoid errors due to lack of convergence we iterate until two
following iterations give identical results up to machine precision. In a few cases the
iterative algorithm ends up oscillating between two or more essentially identical sets of values,
and to cope with this we include the extra stopping criterion that at most $10\, 000$ iterations
are to be run. In practice the iterative algorithm converges in much less than $10\, 000$ iterations.
After having produced $Q_i^{(0)},i=1,\ldots,s$, we use each $Q_i^{(0)}$ as the initial state
in a Markov chain and run $r=3m$ steps with the transition kernel
$P_{\mbox{\tiny ru}}(\cdot|\cdot)$ described in the end of Section \ref{sec:GW}. Note that this
means that on average each maximal clique is visited and updated three times. 
We let $Q_i^{(\ell )}$ denote the state after $\ell=1,\dots,r$ updates, so in
particular $Q_i^{(r)}$ is the resulting final state of the chain.
Letting $x_i^{(0)}$ and $x_i^{(r)}$ denote vectors with the values of the free
variables in $Q_i^{(0)}$ and $Q_i^{(r)}$, respectively,
we have the situation discussed in Section \ref{sec:test}. Before performing a formal hypothesis test we
in the next section present some preliminary exploratory analysis of the simulated $Q_i^{(\ell)}$ values.

\subsection{\label{sec:prelim}Preliminary exploritary analysis}
\begin{figure}
  \vspace*{-2.0cm}
  \begin{center}
    \begin{tabular}{@{}c@{}c@{}c@{}c@{}}
      & \tiny $Q_{24}$ & \tiny $\ln(tr(Q))$ & \tiny $\ln(|Q|)$ \\[-0.3cm]
      \raisebox{1.1cm}{\rotatebox{90}{\begin{tabular}{c}\tiny Graph (a)\\[-0.2cm] \tiny empirical mean\end{tabular}}}&
      \includegraphics[width=4.3cm]{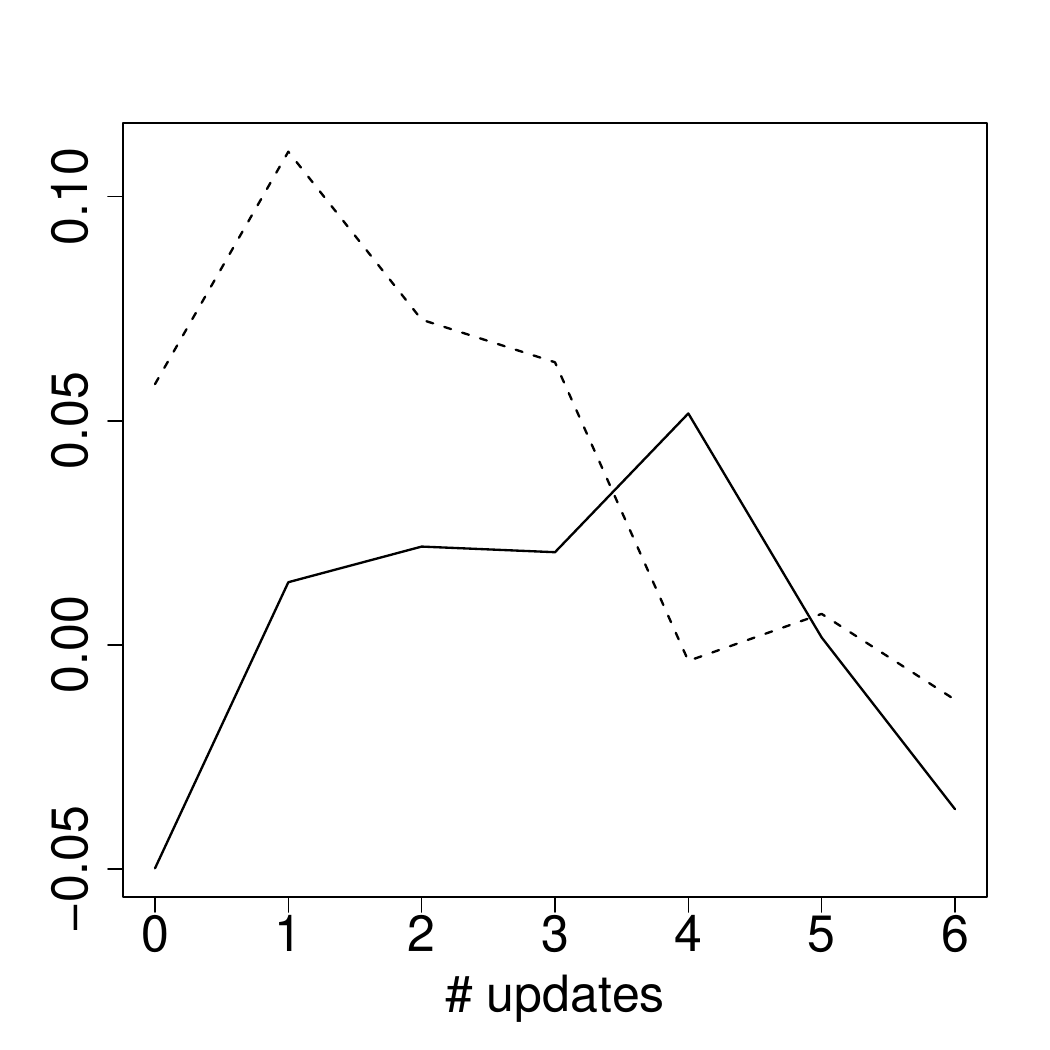} &
      \includegraphics[width=4.3cm]{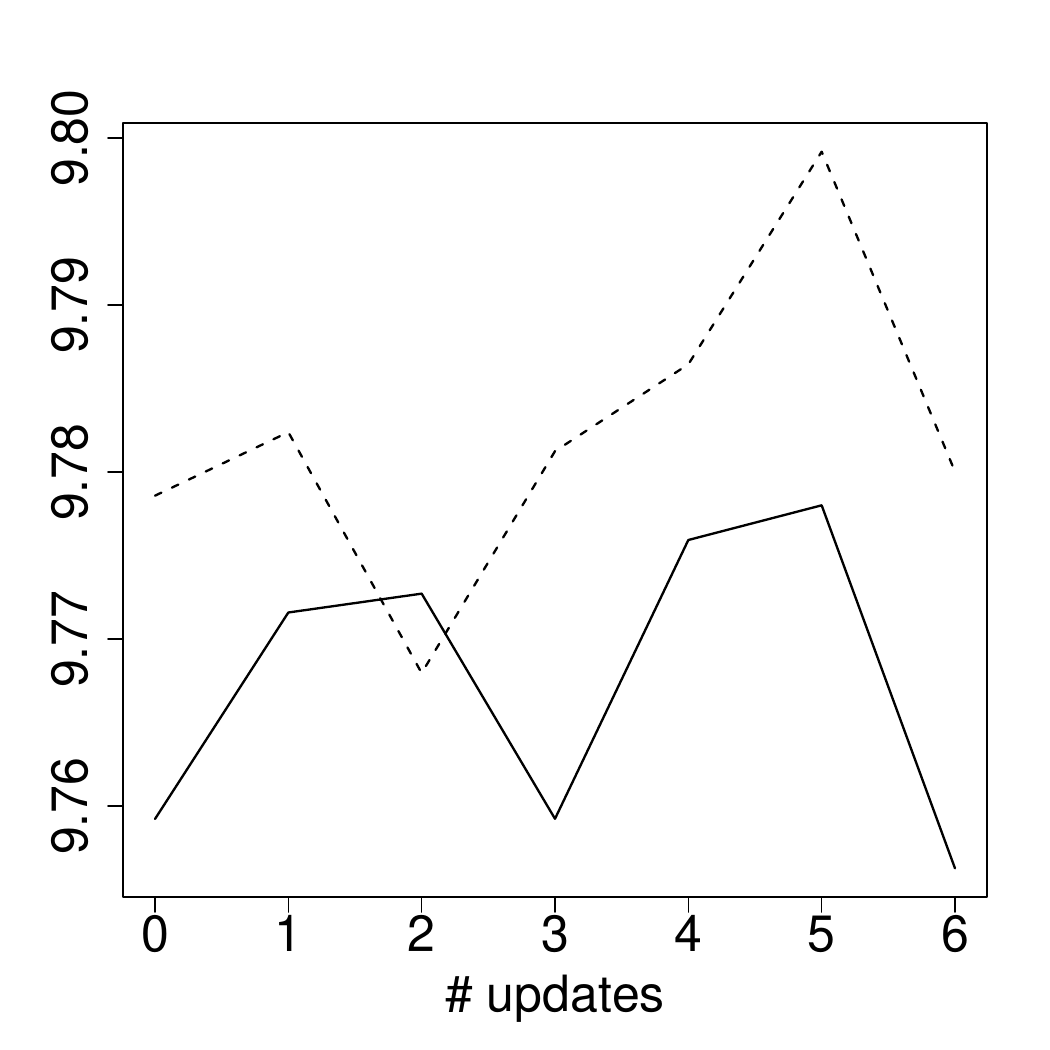} &
      \includegraphics[width=4.3cm]{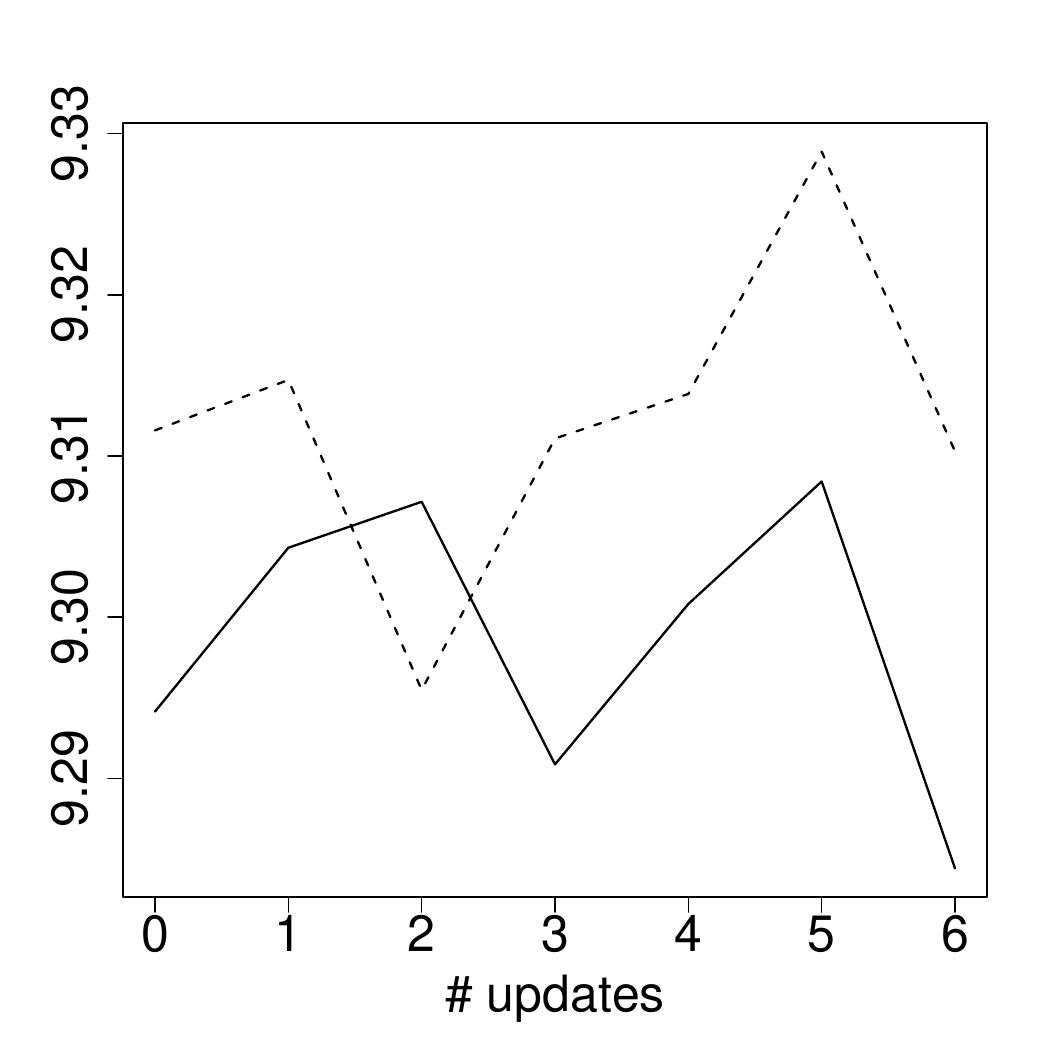} \\[-0.1cm]
      \raisebox{0.3cm}{\rotatebox{90}{\begin{tabular}{c}\tiny Graph (a)\\[-0.2cm] \tiny empirical standard deviation\end{tabular}}}&
      \includegraphics[width=4.3cm]{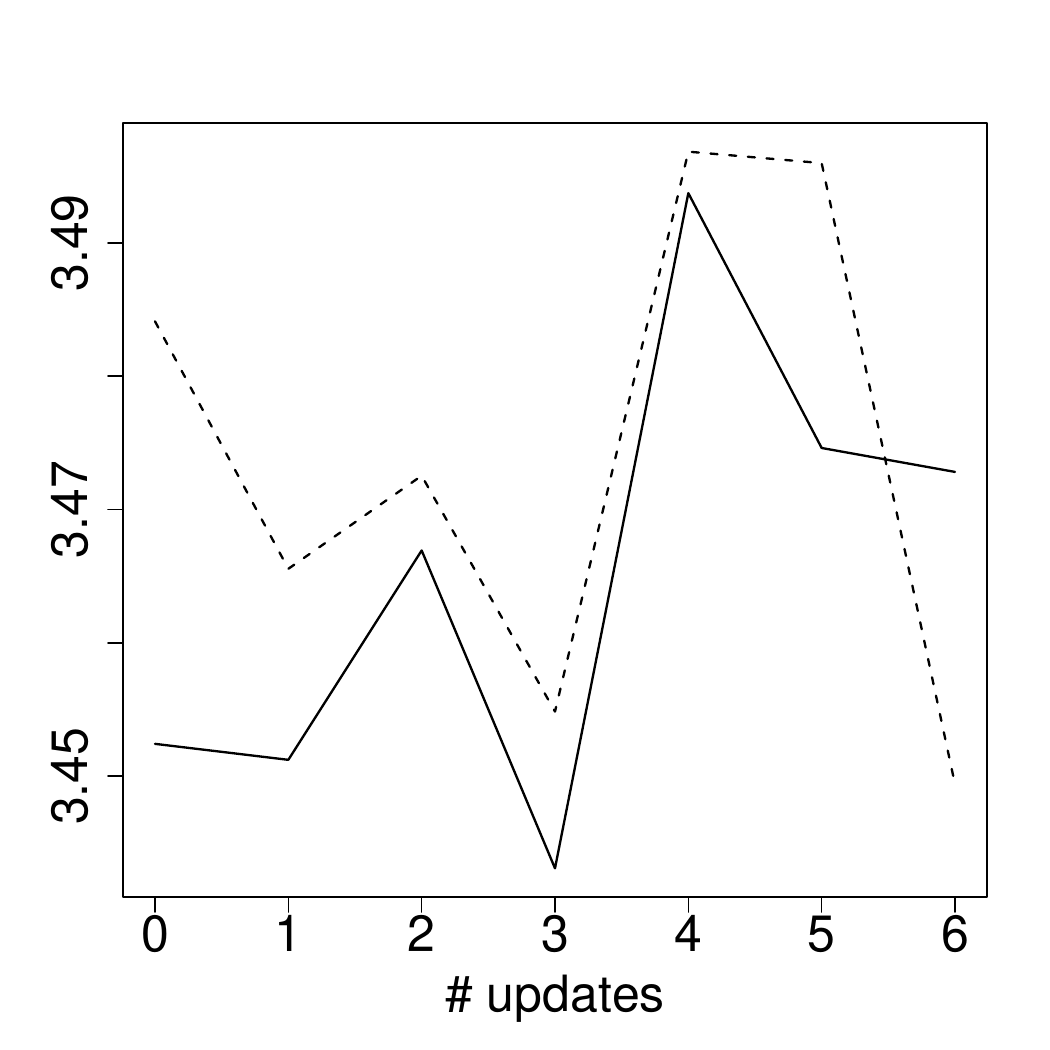} &
      \includegraphics[width=4.3cm]{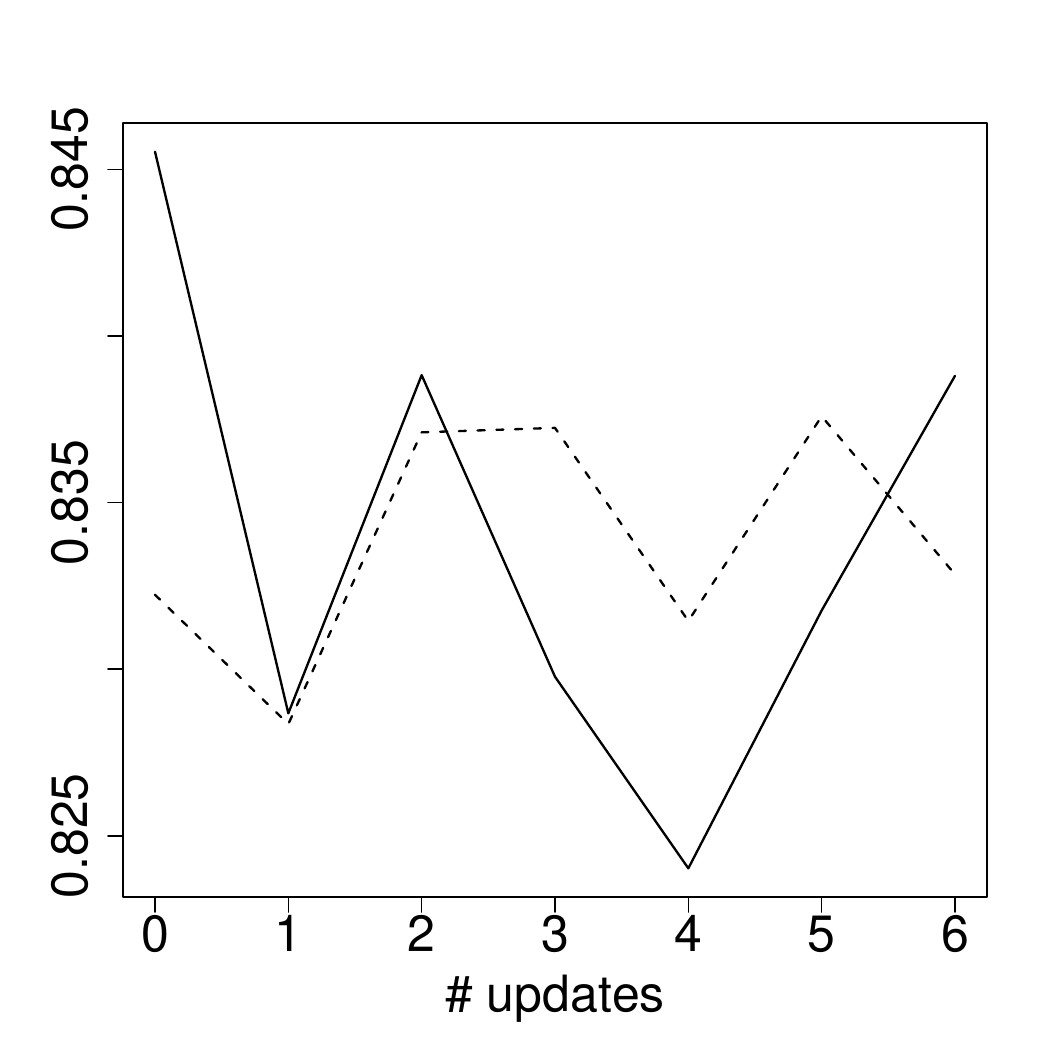} &
      \includegraphics[width=4.3cm]{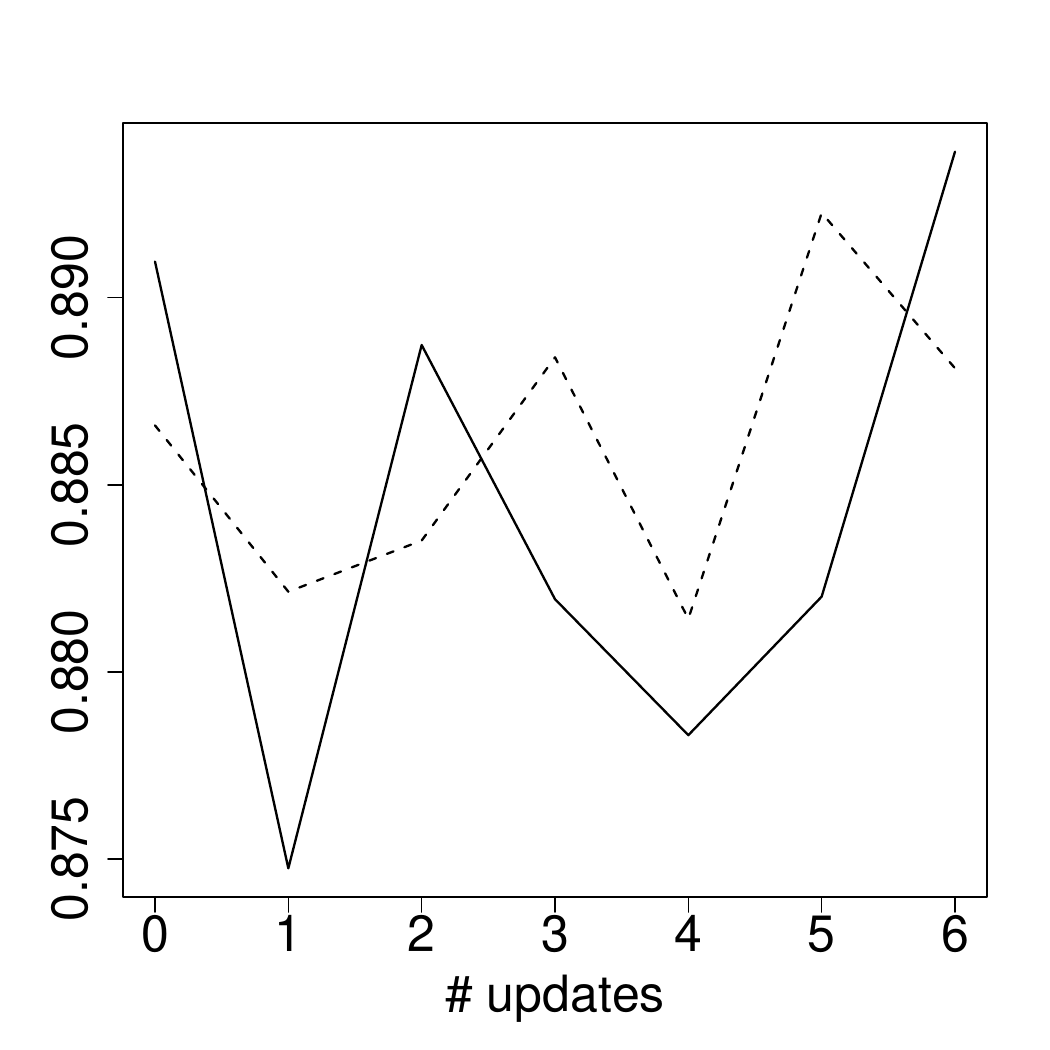} \\[-0.1cm]
      \raisebox{1.1cm}{\rotatebox{90}{\begin{tabular}{c}\tiny Graph (b)\\[-0.2cm] \tiny empirical mean\end{tabular}}}&
      \includegraphics[width=4.3cm]{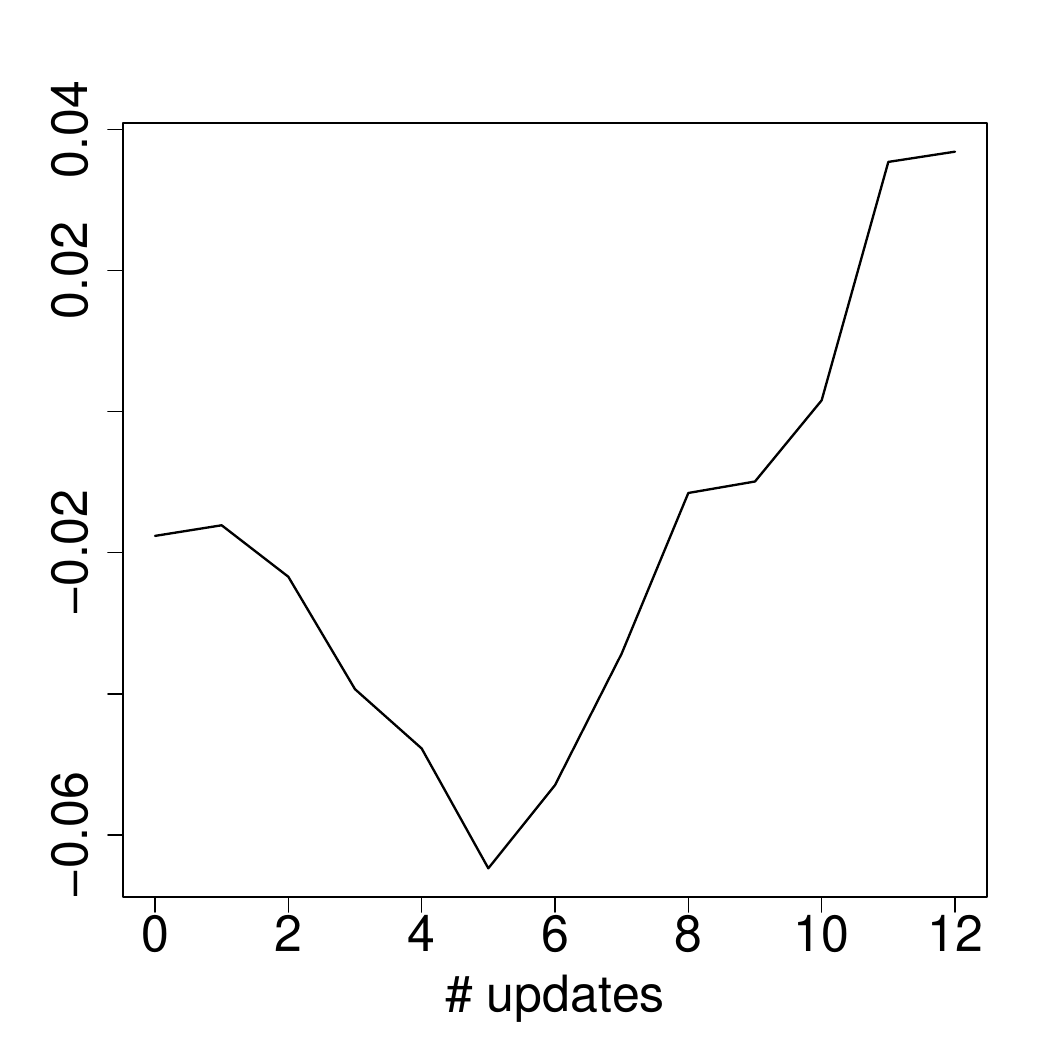} &
      \includegraphics[width=4.3cm]{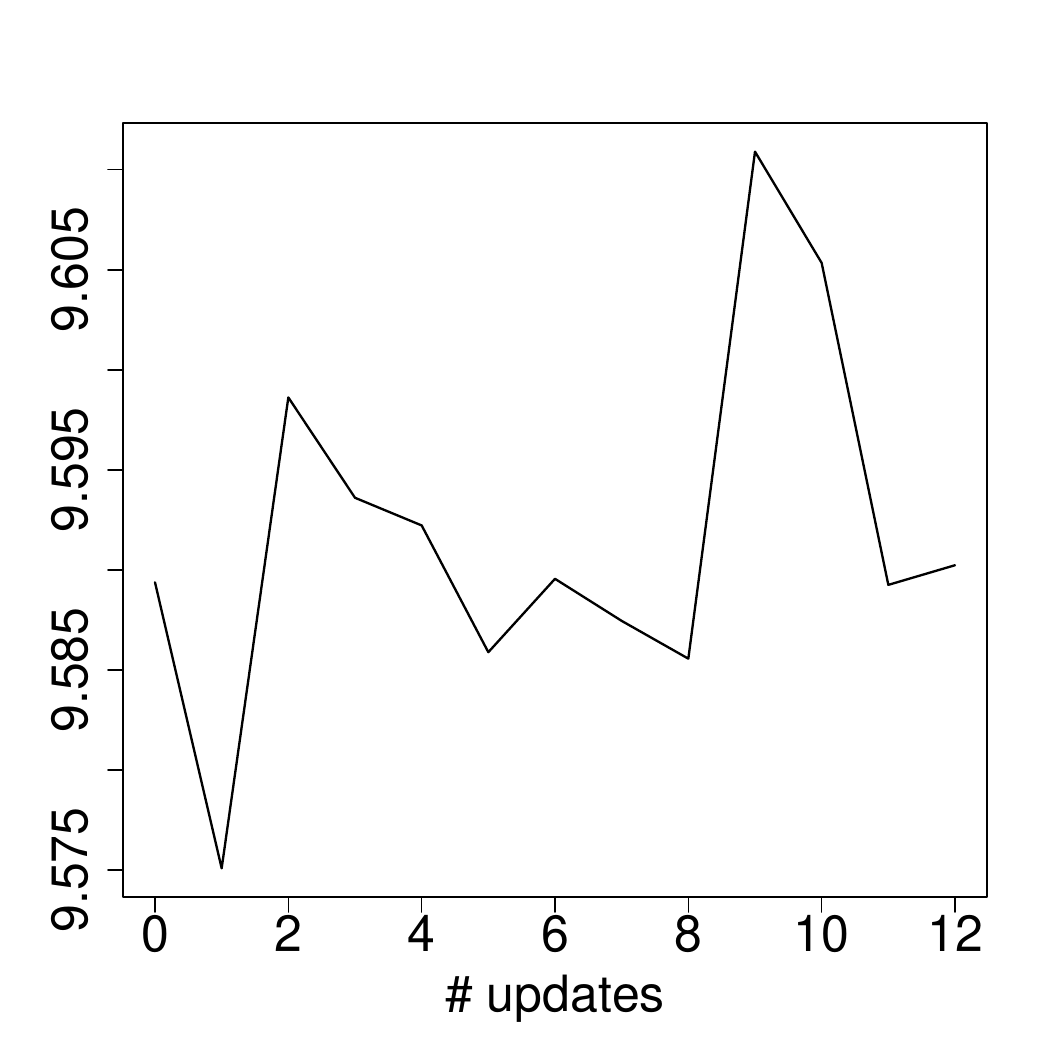} &
      \includegraphics[width=4.3cm]{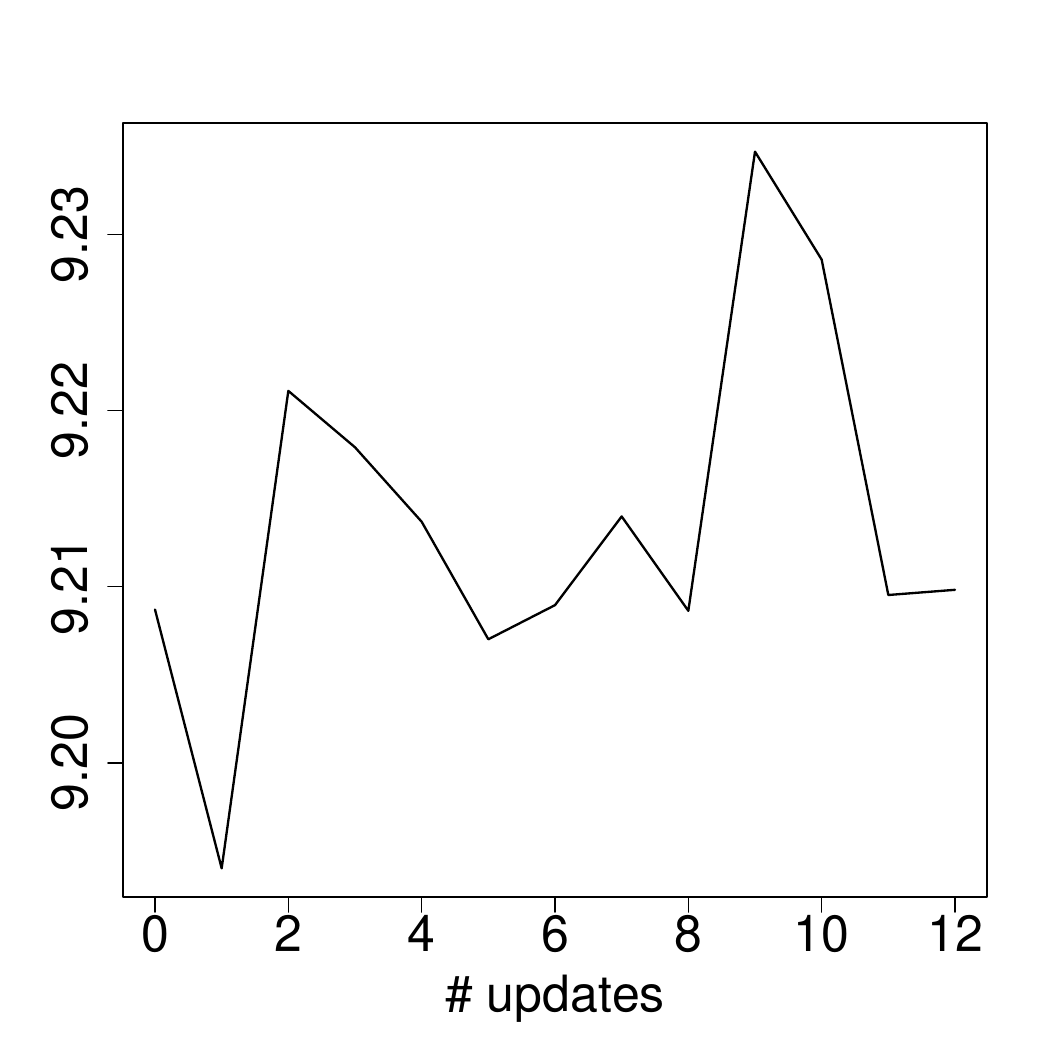} \\[-0.1cm]
      \raisebox{0.3cm}{\rotatebox{90}{\begin{tabular}{c}\tiny Graph (b)\\[-0.2cm] \tiny empirical standard deviation\end{tabular}}}&
      \includegraphics[width=4.3cm]{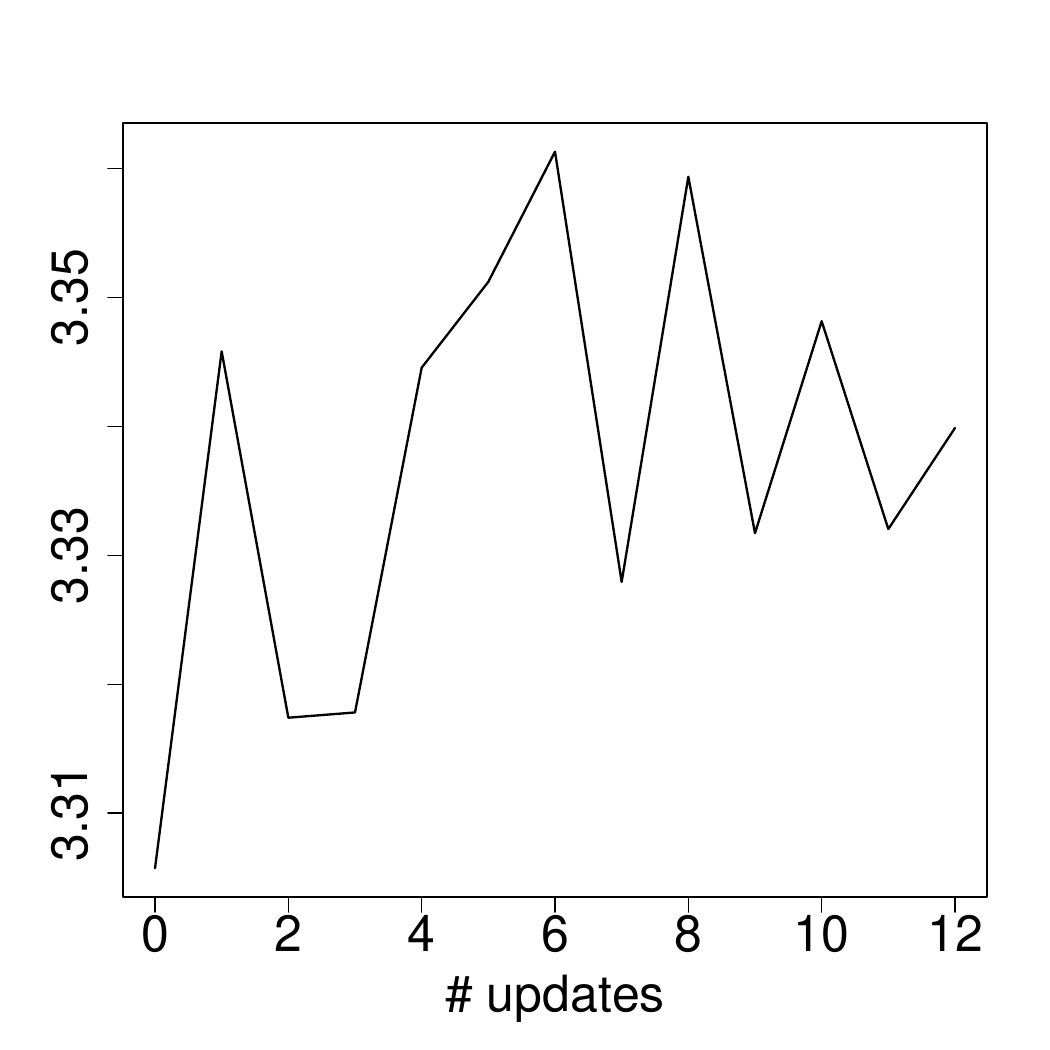} &
      \includegraphics[width=4.3cm]{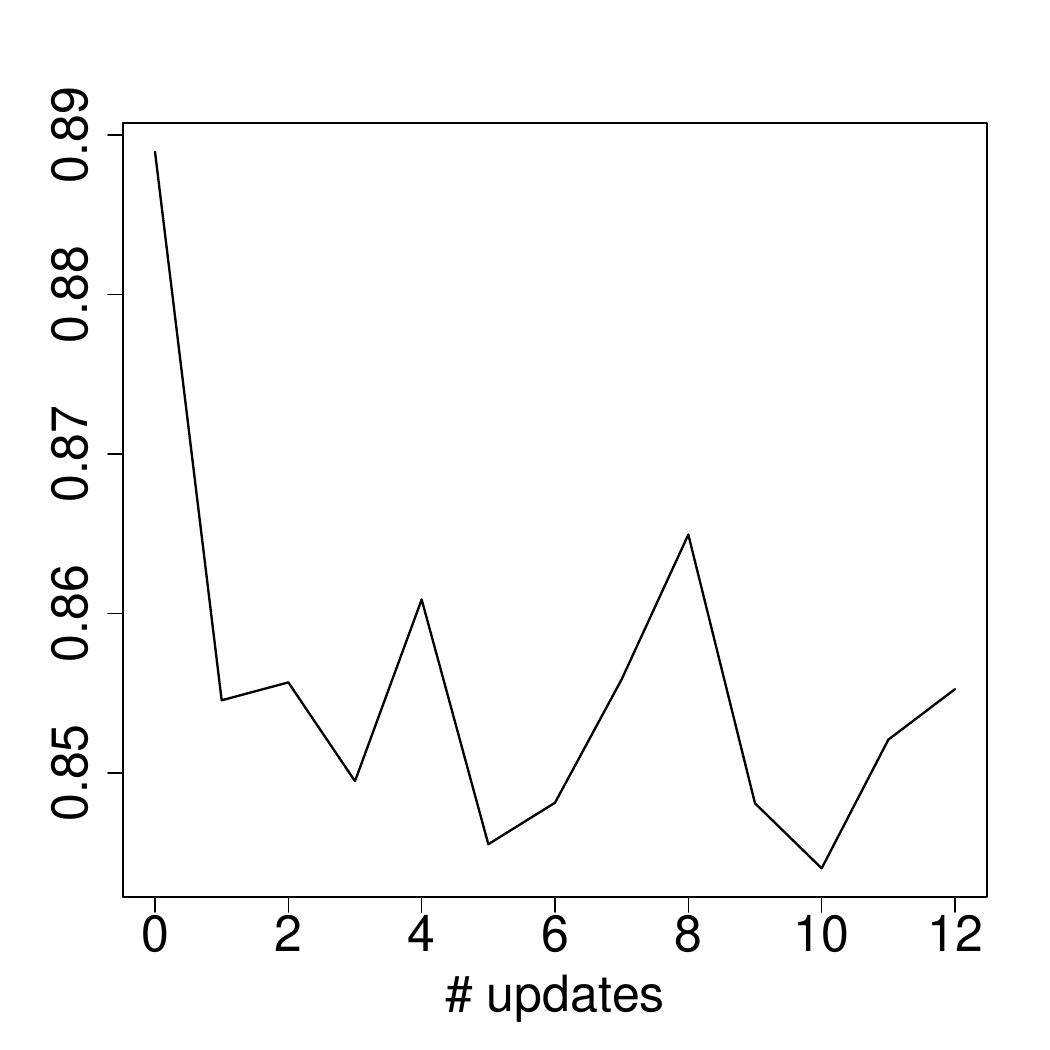} &
      \includegraphics[width=4.3cm]{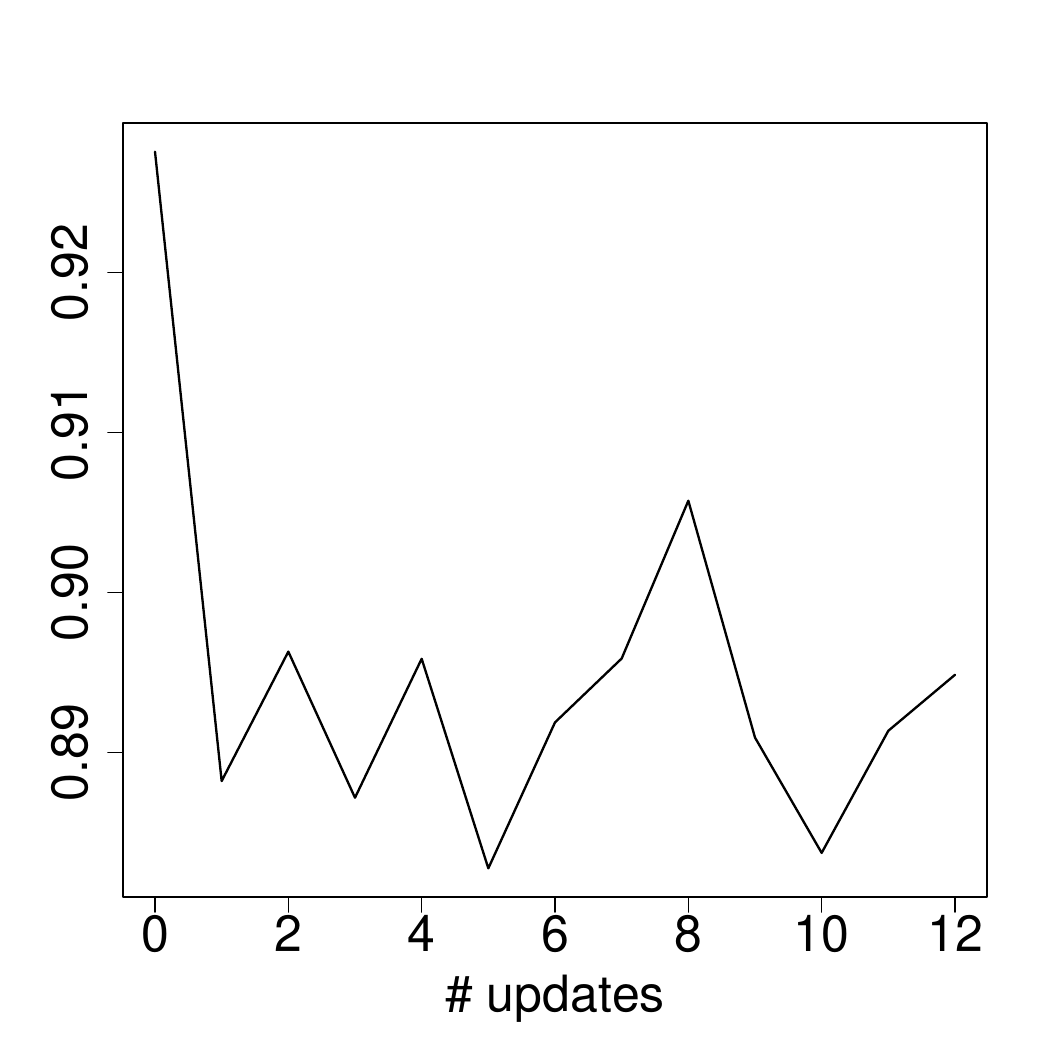} \\[-0.1cm]
      
    \end{tabular}
  \end{center}
  \caption{\label{fig:caseABTrace}Trace plots of scalar functions of the
    simulation results for the two $p=4$ node graph models.
    In the left column
    the scalar value used is the value of element $(2,4)$ of the simulated matrix $Q$, the middle
    column is for the logarithm of the trace of $Q$, and the right column is for the logarithm
    of the determinant of $Q$. For the model based on the graph shown in Figure \ref{fig:graph}(a),
    the two upper rows show the empirical means and the empirical standard deviations, respectively,
    over the $s$ simulated chains. The solid and dashed curves show the results when initialising with the claimed
    sampler and the exact sampler,
    respectively. For the model based on the graph shown in Figure \ref{fig:graph}(b),
    the two lower rows correspondingly show the empirical means and standard deviations over the $s$
    simulated chains. As this graph is not decomposable, only results for the claimed general 
    sampler is included.}
\end{figure}

Figure \ref{fig:caseABTrace}
shows trace plots based on the simulated $Q_i^{(\ell)}$ values when using
the two $p=4$ node graph models shown in Figures \ref{fig:graph}(a) and (b). 
The number of updates $\ell=0,\ldots,r$ is along
the x-axis and scalar functions of the simulated $Q_i^{(\ell)}$ are along the y-axis.
The two upper rows show results for the model based on the graph in Figure \ref{fig:graph}(a), and
the two lower rows show corresponding results for the graph in Figure \ref{fig:graph}(b).
From left to right, the three plots in the upper row show the empirical means
of element $(2,4)$ of $Q_i^{(\ell)}$, $\ln(\mbox{tr}(Q_i^{(\ell)}))$ and $\ln |Q_i^{(\ell)}|$, respectively.
The solid and dashed curves are based on the results when initialising with the claimed sampler and the exact sampler,
respectively. The second row
shows empirical standard deviations for the same three scalar functions of the simulated matrices.
One can note that it is not possible to identify any burn-in period from the plots in these two upper
rows. The two lower rows show corresponding results for the model based on the graph in Figure \ref{fig:graph}(b),
except that here only results for the claimed sampler is included since the exact 
sampler is not valid for non-decomposable graphs.
From the empirical mean results in the third row it is again not possible to identify any burn-in period.
However, from the estimated standard deviations in the bottom row we see a clear decrease in the standard
deviations of $\ln(\mbox{tr}Q_i^{(\ell)})$ and $\ln|Q_i^{(\ell)}|$ from $\ell=0$ to $\ell=1$. This is a first
indication that
the claimed general sampler is perhaps not correctly simulating
from the specified G-Wishart distribution.

\begin{figure}
  \vspace*{-2.0cm}
  \begin{center}
    \begin{tabular}{@{}c@{}c@{}c@{}c@{}}
      & \tiny $Q_{24}$ & \tiny $\ln(tr(Q))$ & \tiny $\ln(|Q|)$ \\[-0.3cm]
      \raisebox{1.1cm}{\rotatebox{90}{\begin{tabular}{c}\tiny Graph (c)\\[-0.2cm] \tiny empirical mean\end{tabular}}}&
      \includegraphics[width=4.3cm]{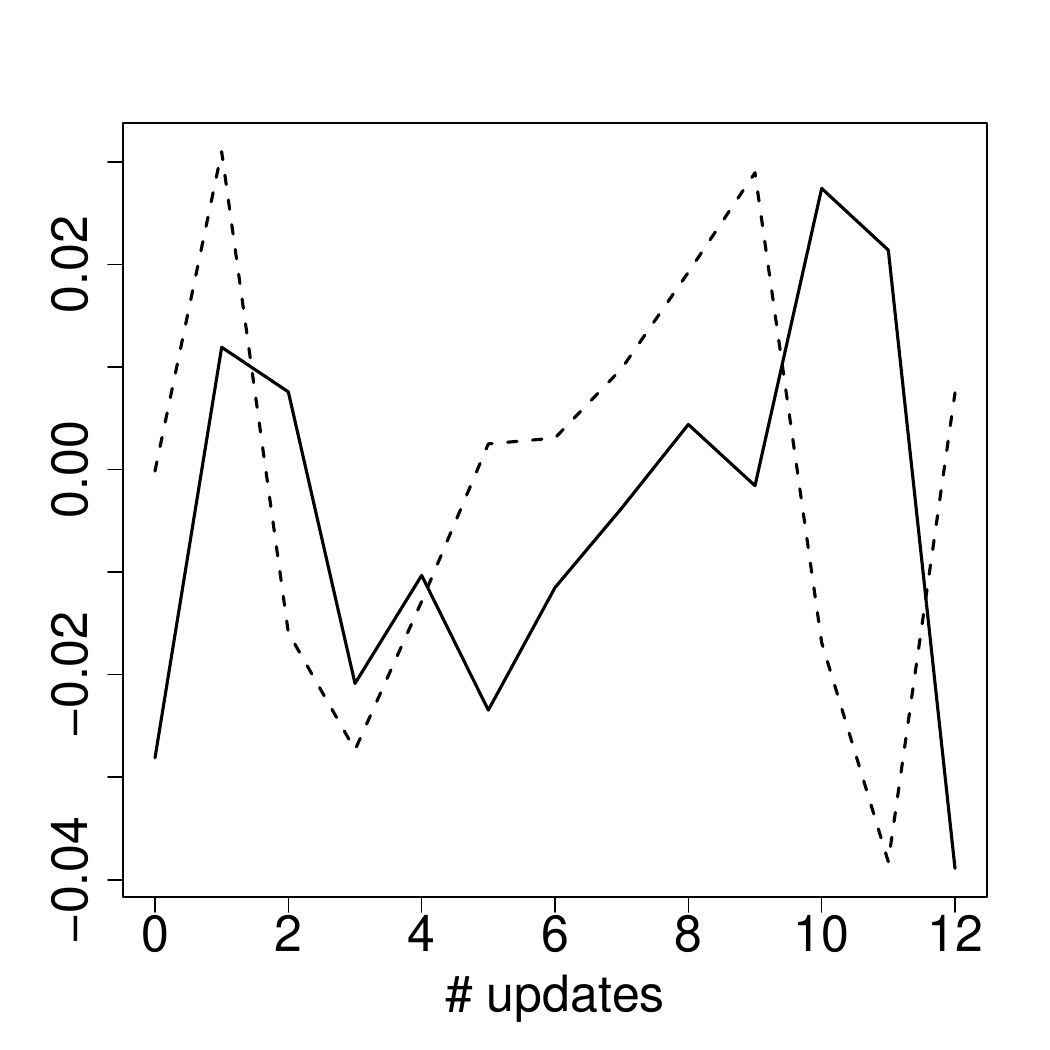} &
      \includegraphics[width=4.3cm]{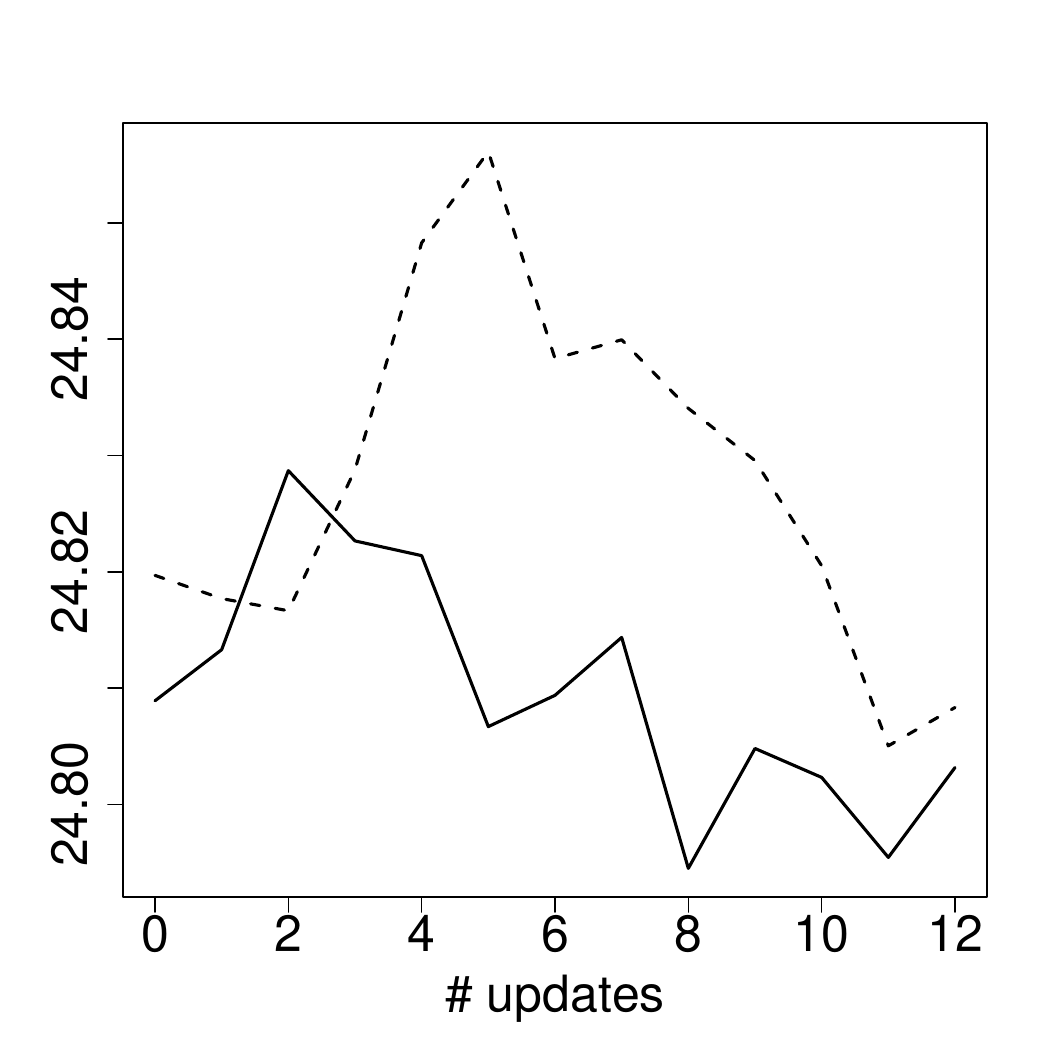} &
      \includegraphics[width=4.3cm]{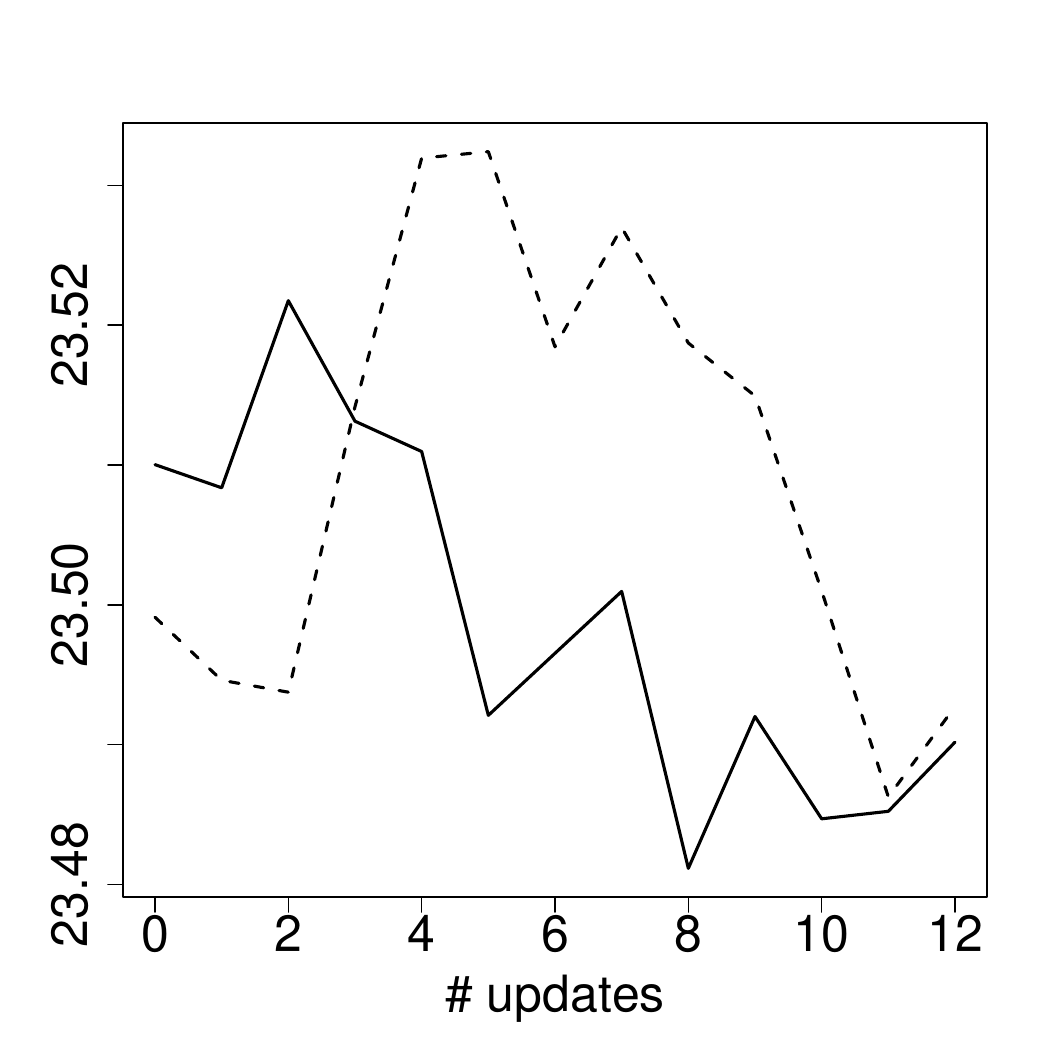} \\[-0.1cm]
      \raisebox{0.3cm}{\rotatebox{90}{\begin{tabular}{c}\tiny Graph (c)\\[-0.2cm] \tiny empirical standard deviation\end{tabular}}}&
      \includegraphics[width=4.3cm]{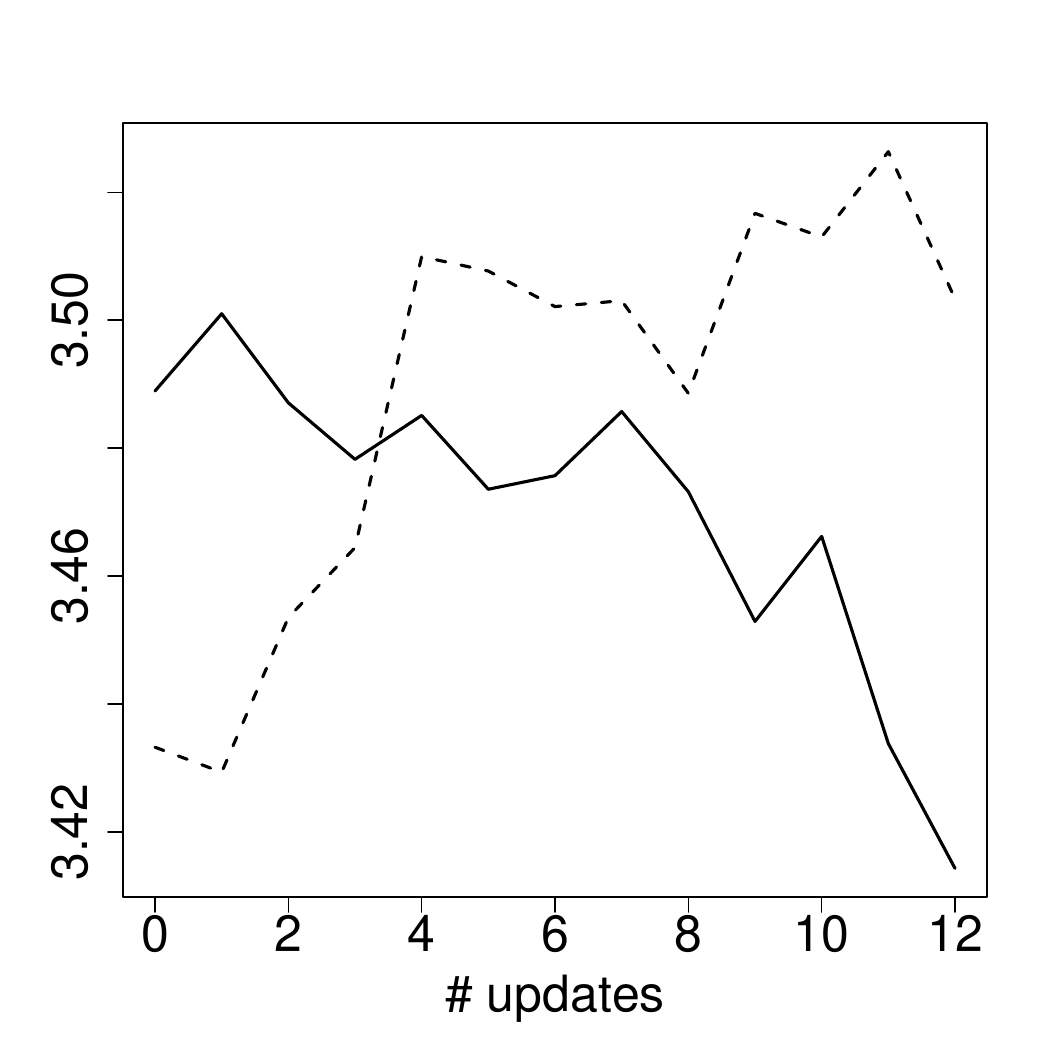} &
      \includegraphics[width=4.3cm]{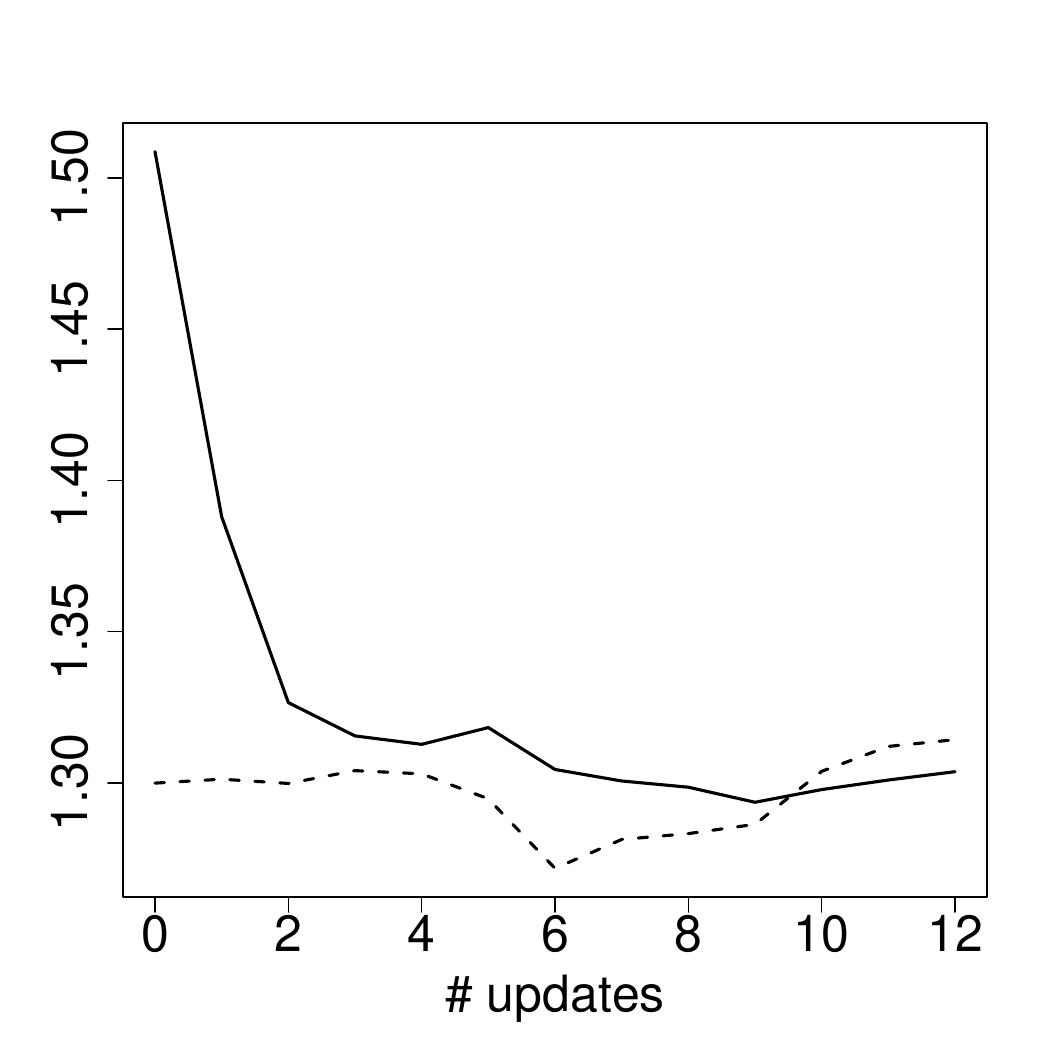} &
      \includegraphics[width=4.3cm]{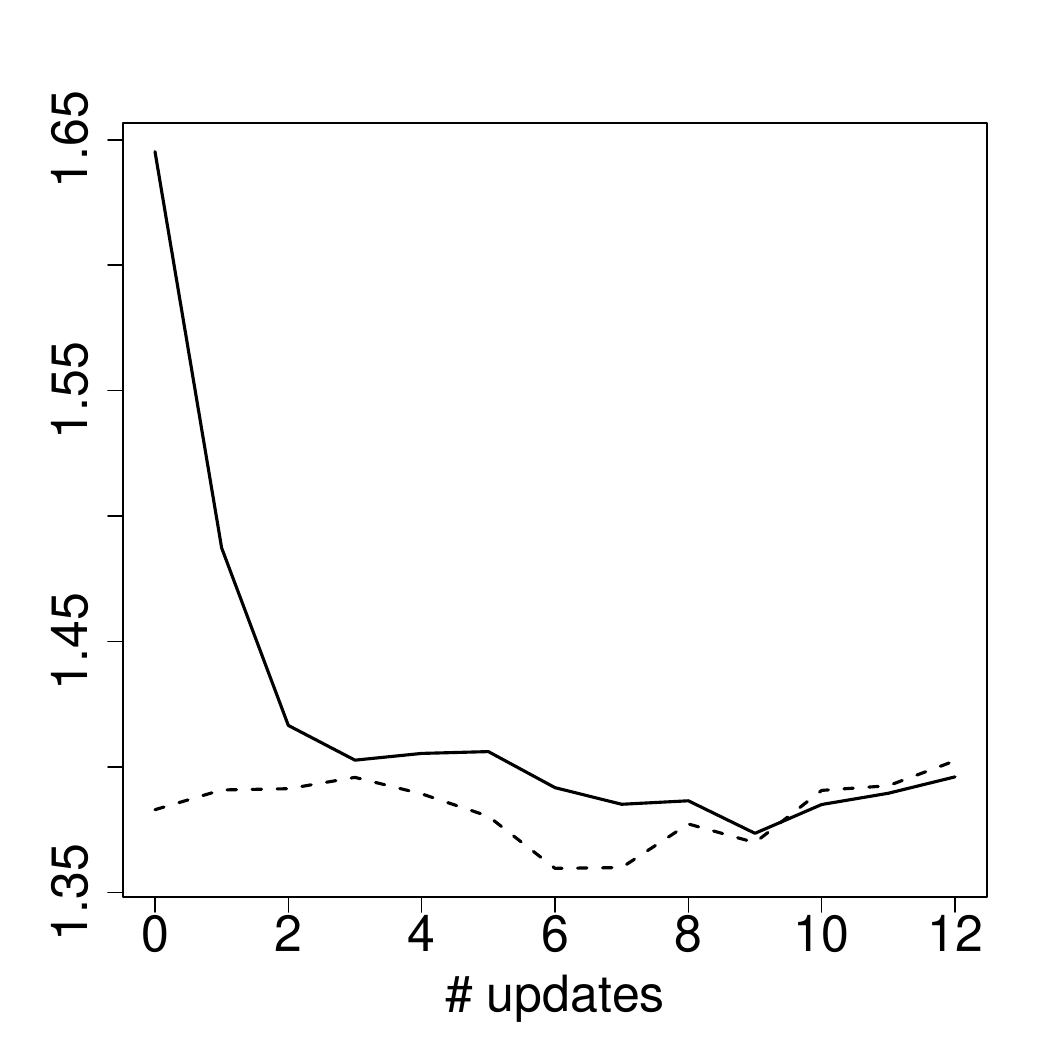} \\[-0.1cm]
      \raisebox{1.1cm}{\rotatebox{90}{\begin{tabular}{c}\tiny Graph (d)\\[-0.2cm] \tiny empirical mean\end{tabular}}}&
      \includegraphics[width=4.3cm]{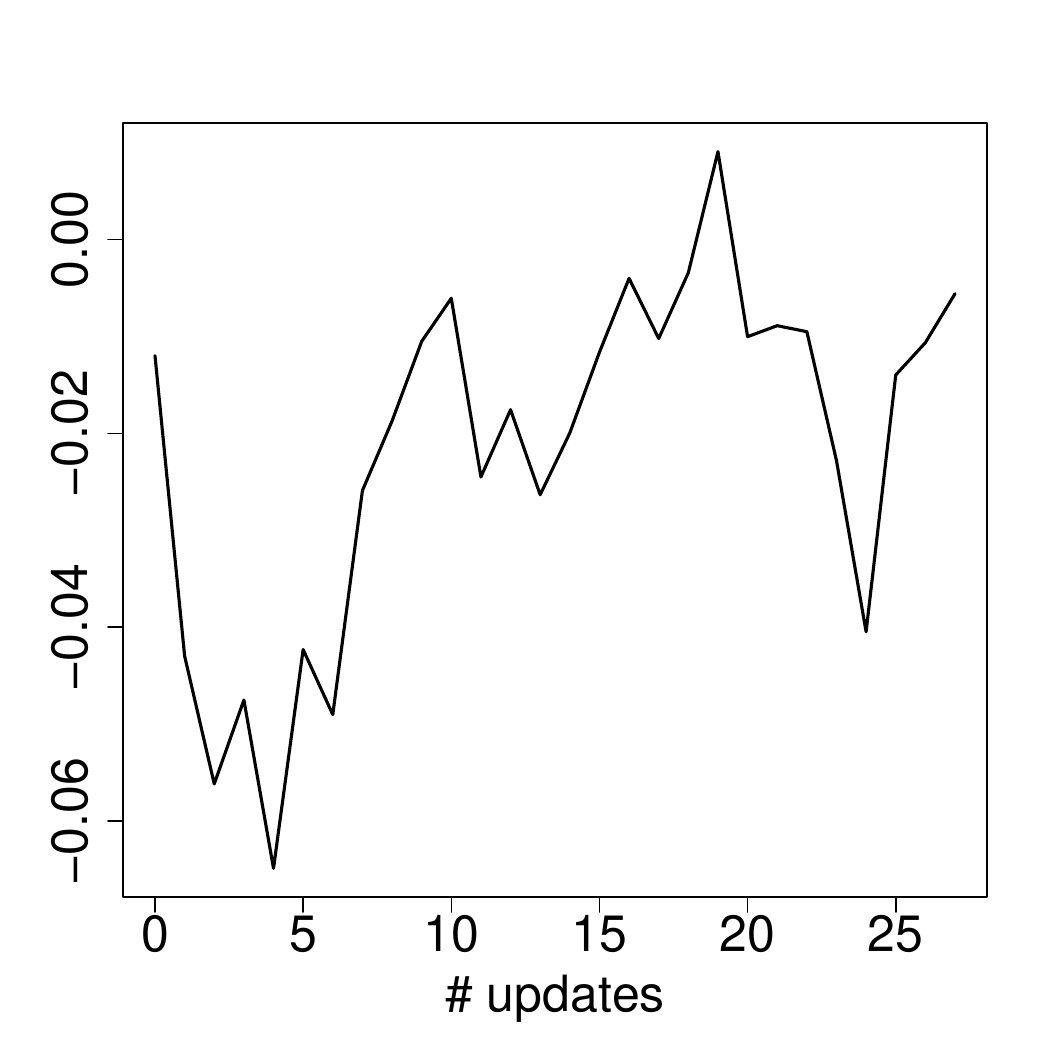} &
      \includegraphics[width=4.3cm]{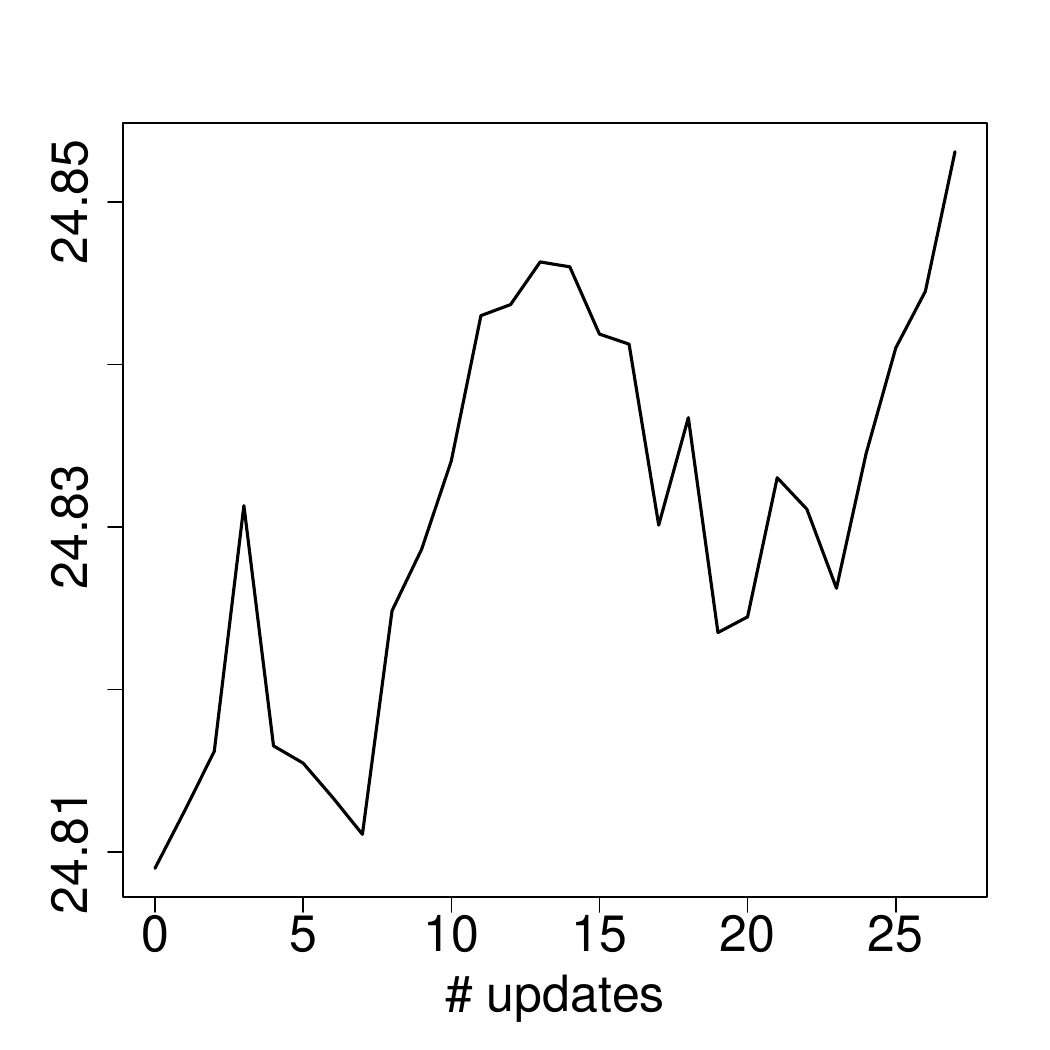} &
      \includegraphics[width=4.3cm]{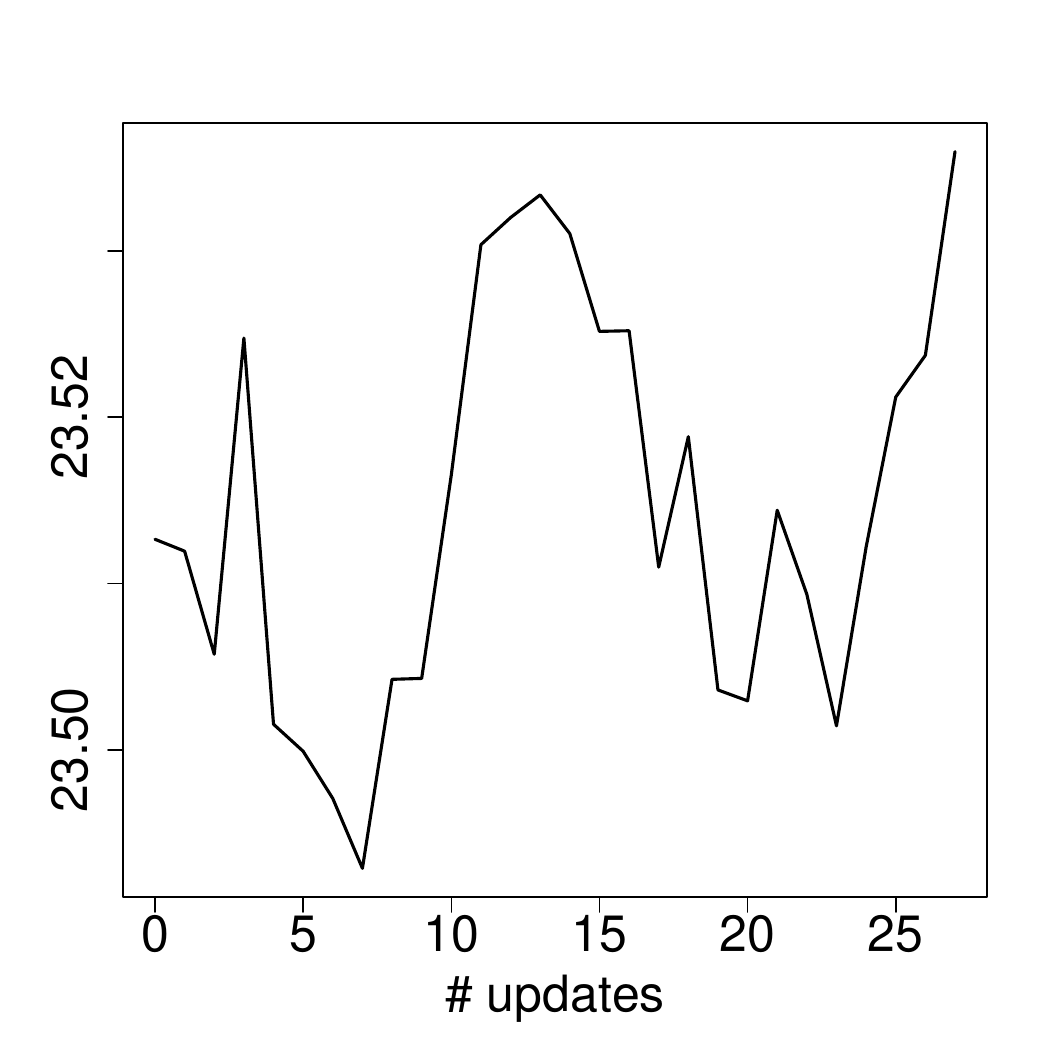} \\[-0.1cm]
      \raisebox{0.3cm}{\rotatebox{90}{\begin{tabular}{c}\tiny Graph (d)\\[-0.2cm] \tiny empirical standard deviation\end{tabular}}}&
      \includegraphics[width=4.3cm]{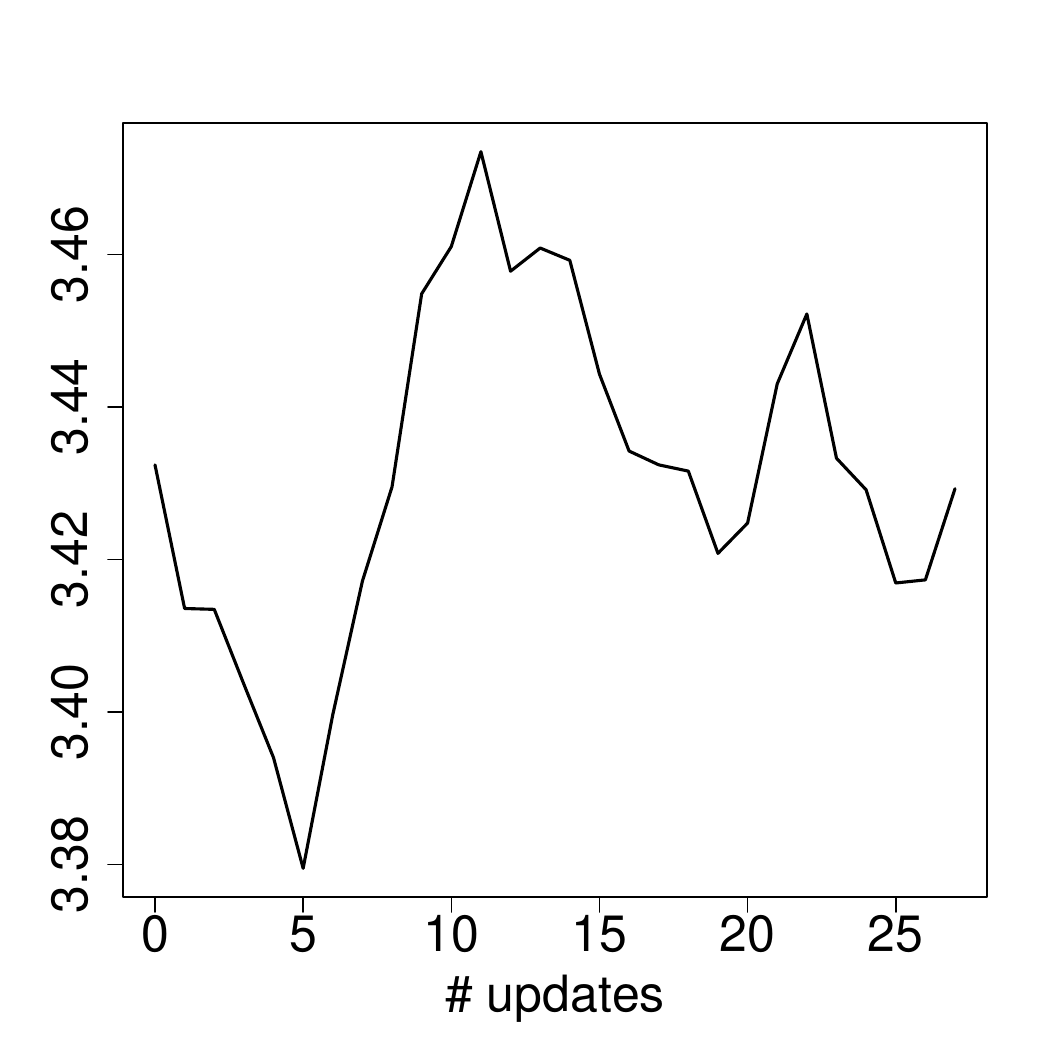} &
      \includegraphics[width=4.3cm]{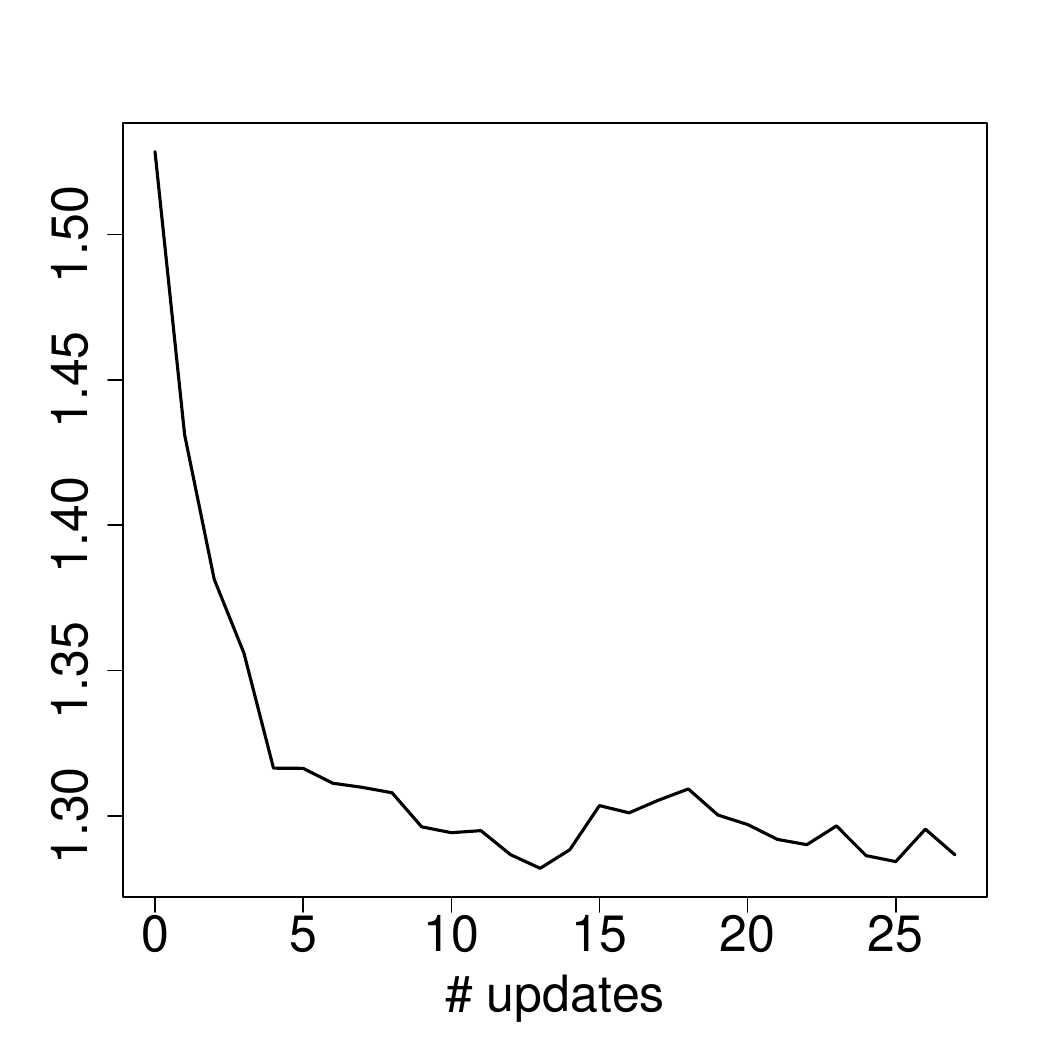} &
      \includegraphics[width=4.3cm]{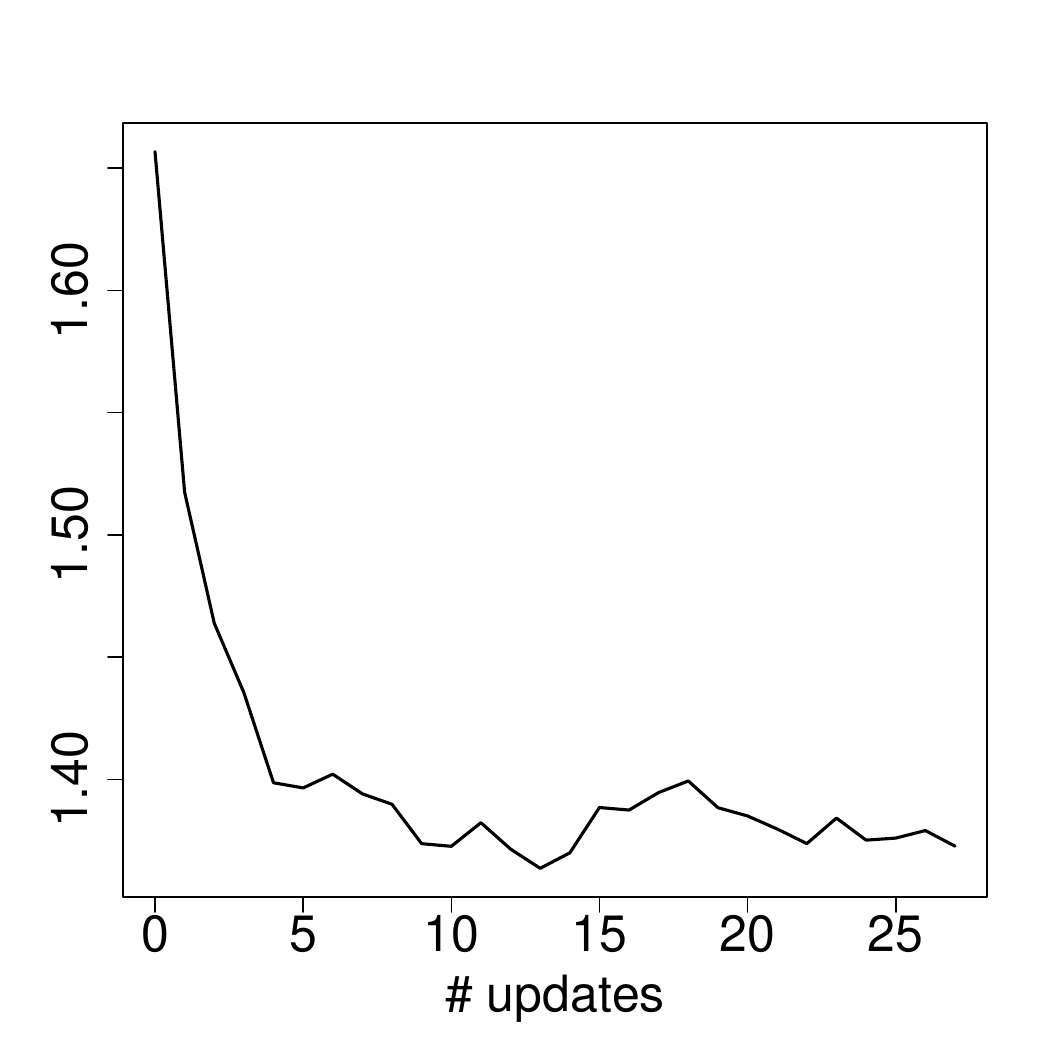} \\[-0.1cm]
      
    \end{tabular}
  \end{center}
  \caption{\label{fig:caseCDTrace}Trace plots of scalar functions of the
    simulation results for the two $p=10$ node graph models.
    In the left column
    the scalar value used is the value of element $(2,4)$ of the simulated matrix $Q$, the middle
    column is for the logarithm of the trace of $Q$, and the right column is for the logarithm
    of the determinant of $Q$. For the model based on the graph shown in Figure \ref{fig:graph}(c),
    the two upper rows show the empirical means and the empirical standard deviations, respectively,
    over the $s$ simulated chains. The solid and dashed curves show the results when initialising with the claimed
    sampler and the exact sampler, 
    respectively. For the model based on the graph shown in Figure \ref{fig:graph}(d),
    the two lower rows correspondingly show the empirical means and standard deviations over the $s$
    simulated chains. As this graph is not decomposable, only results for the claimed general
    sampler is included.}
\end{figure}

Figure \ref{fig:caseCDTrace} shows the exact same types of trace plots as Figure \ref{fig:caseABTrace}, but now for
the two $p=10$ node graph models.
Also for this model it is not possible
to identify any clear burn-in effect from the plots in the left column or for the empirical means.
However, from the estimated standard deviations for $\ln(\mbox{tr}(Q_i^{(\ell)}))$ and
$\ln|Q_i^{(\ell)}|$ we see a clear decrease as $\ell$ increase from zero when using the claimed general sampler.

\begin{figure}
  \vspace*{-1.5cm}
  \begin{center}
    \begin{tabular}{@{}c@{}c@{}c@{}}
      \mbox{\tiny Graph (a)} &
      \mbox{\tiny Graph (a)} &
      \mbox{\tiny Graph (a)} \\[-0.4cm]
      \begin{tabular}{@{}c@{}c@{}}
        \raisebox{2.0cm}{\rotatebox{90}{\tiny cdf}}
        \includegraphics[width=4.4cm]{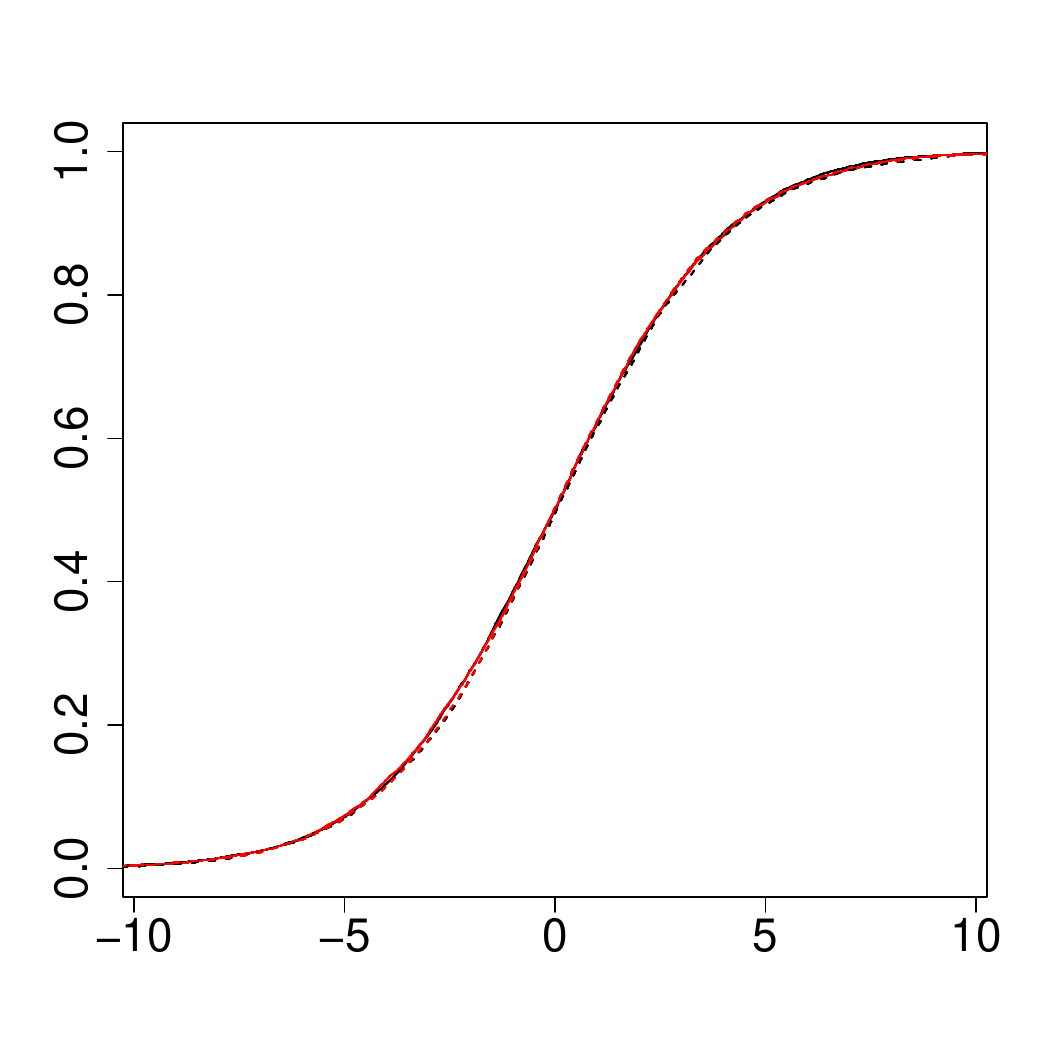}\\[-0.6cm]
        \mbox{\tiny $Q_{24}$}
      \end{tabular}&
      \begin{tabular}{@{}c@{}c@{}}
        \includegraphics[width=4.4cm]{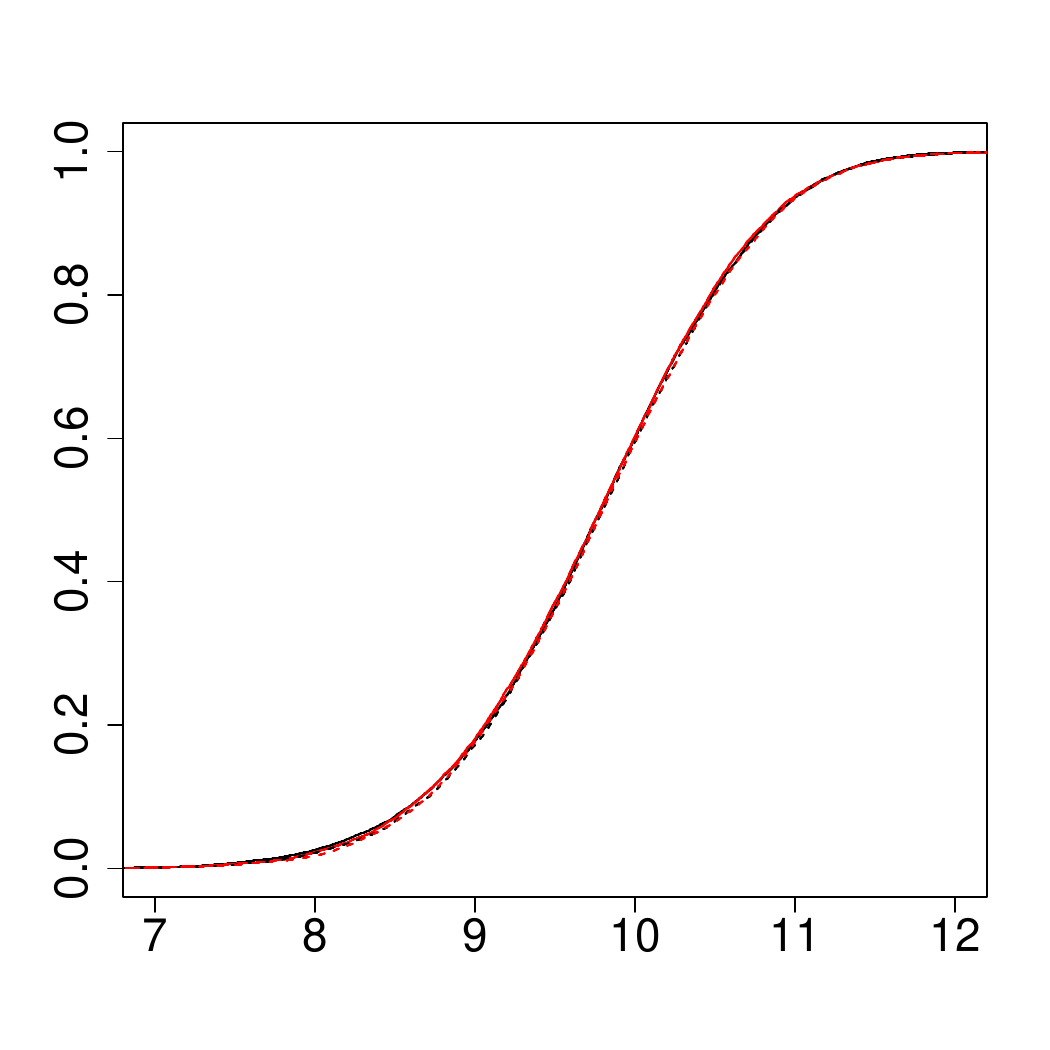}\\[-0.6cm]
        \mbox{\tiny $\ln(\mbox{tr}(Q))$}
      \end{tabular}&
      \begin{tabular}{@{}c@{}c@{}}
        \includegraphics[width=4.4cm]{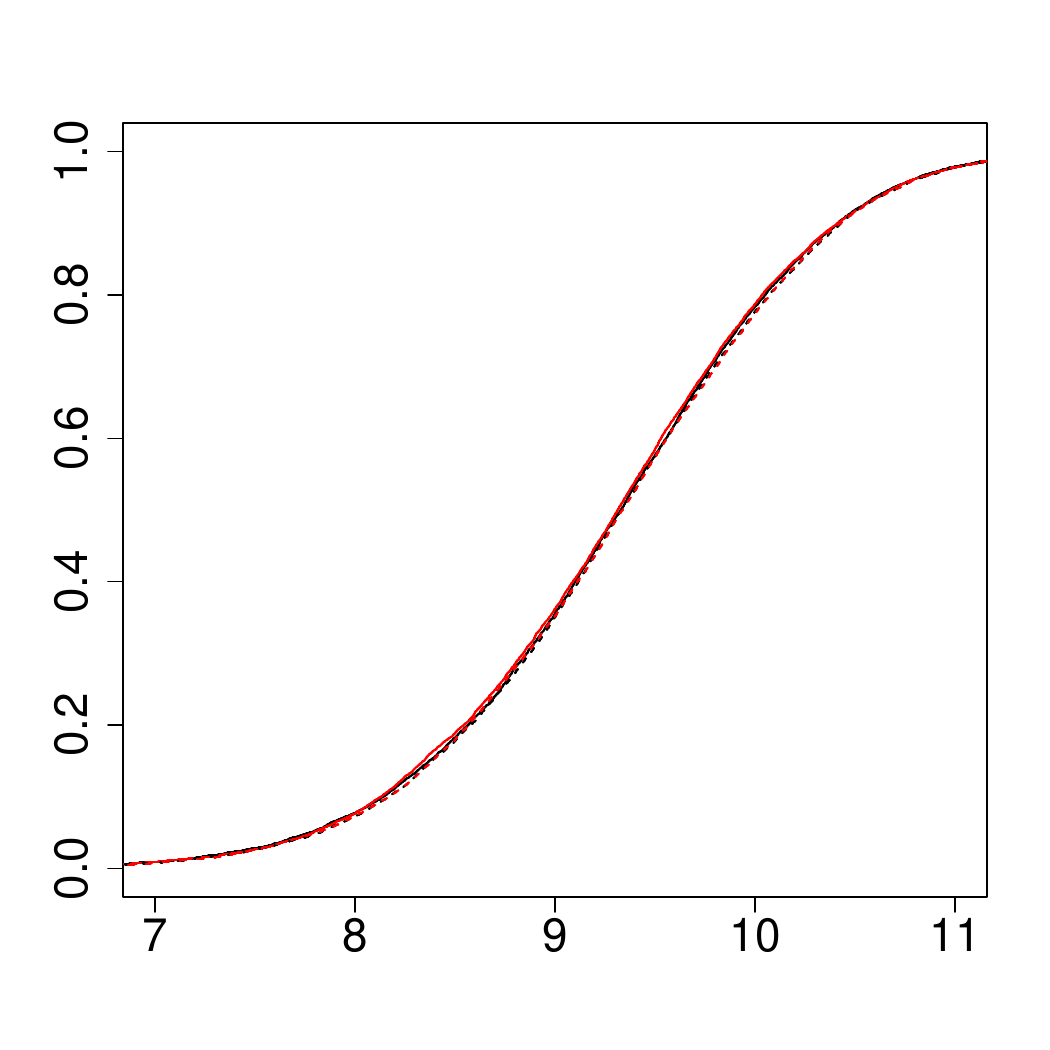}\\[-0.6cm]
        \mbox{\tiny $\ln(tr(Q))$}
      \end{tabular}\\
      \mbox{\tiny Graph (b)} &
      \mbox{\tiny Graph (b)} &
      \mbox{\tiny Graph (b)} \\[-0.4cm]
      \begin{tabular}{@{}c@{}c@{}}
        \raisebox{2.0cm}{\rotatebox{90}{\tiny cdf}}
        \includegraphics[width=4.4cm]{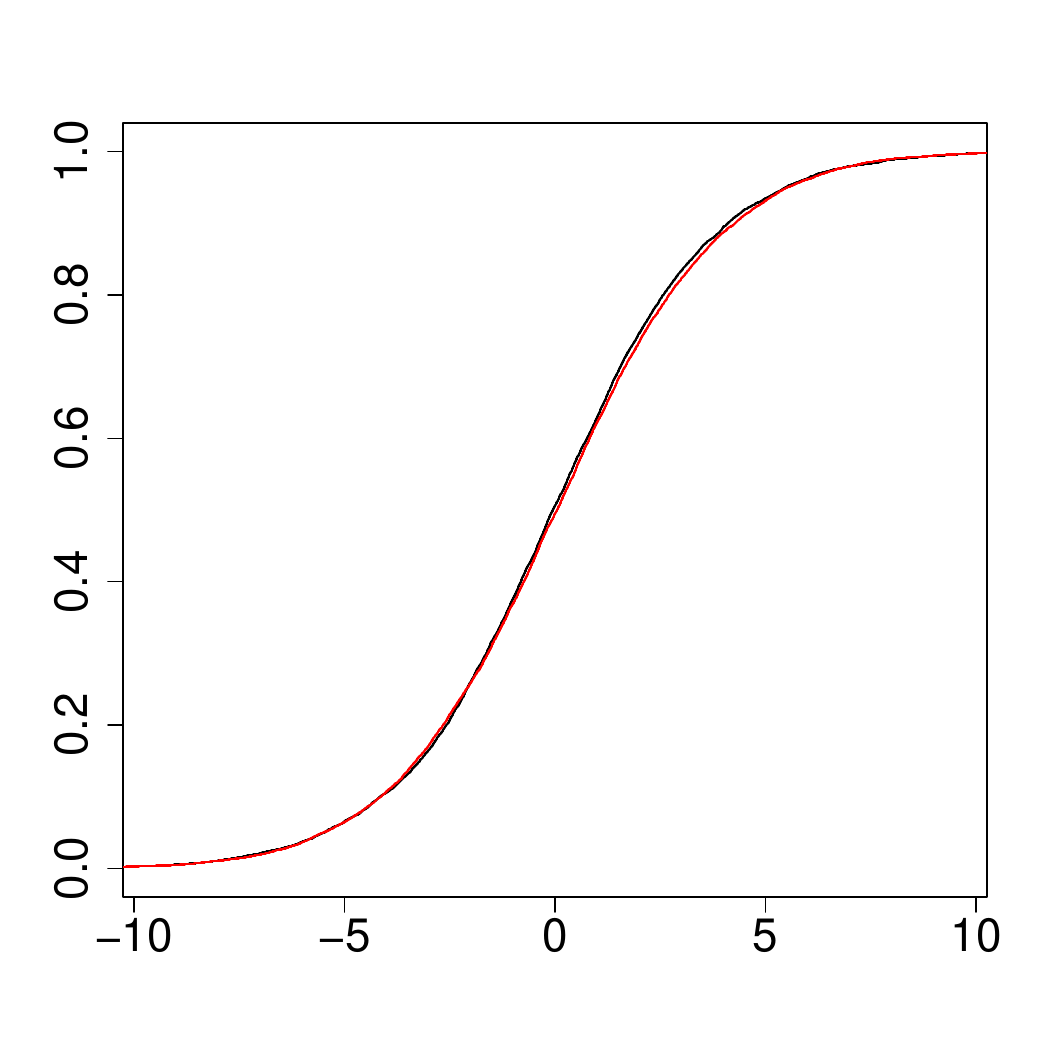}\\[-0.6cm]
        \mbox{\tiny $Q_{24}$}
      \end{tabular} &
      \begin{tabular}{@{}c@{}c@{}}
        \includegraphics[width=4.4cm]{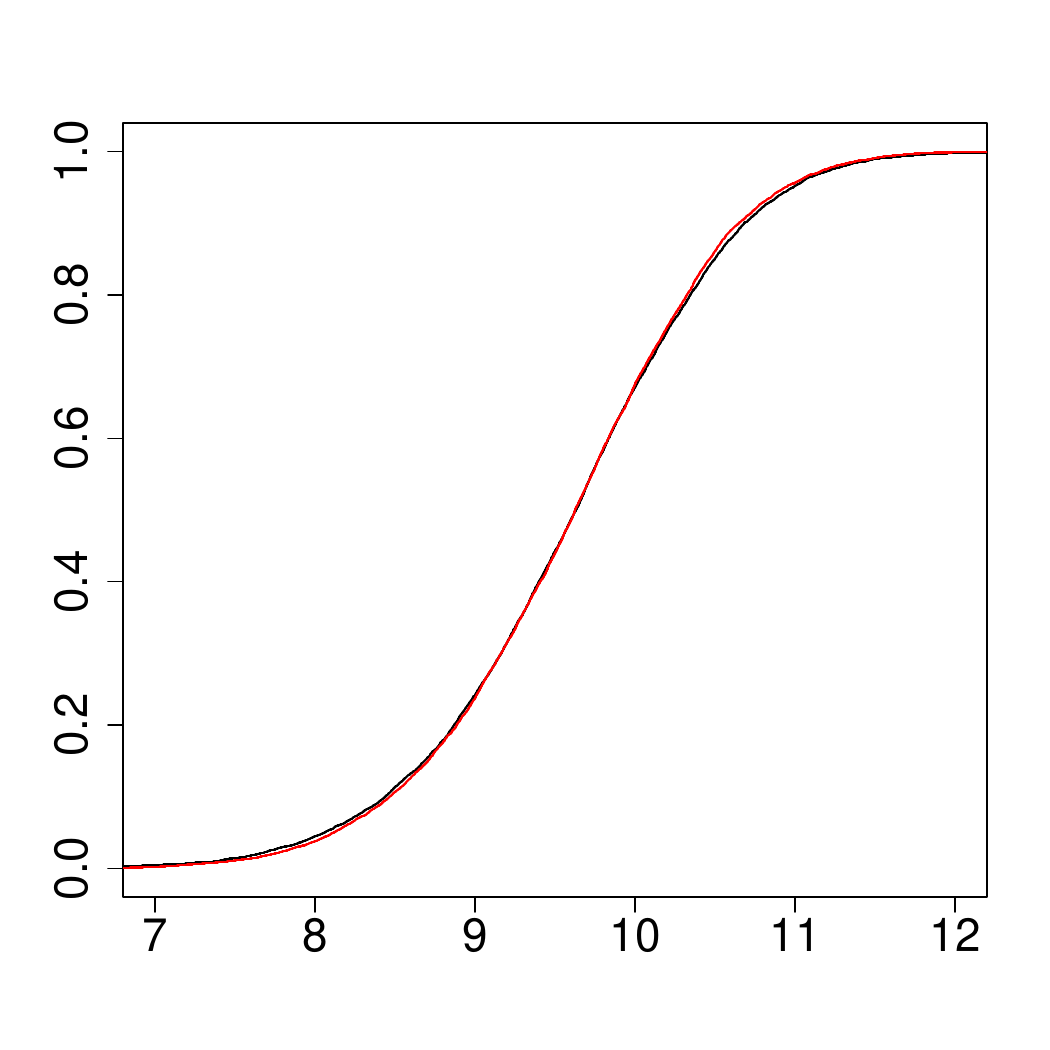}\\[-0.6cm]
        \mbox{\tiny $\ln(\mbox{tr}(Q))$}
      \end{tabular}  &
      \begin{tabular}{@{}c@{}c@{}}
        \includegraphics[width=4.4cm]{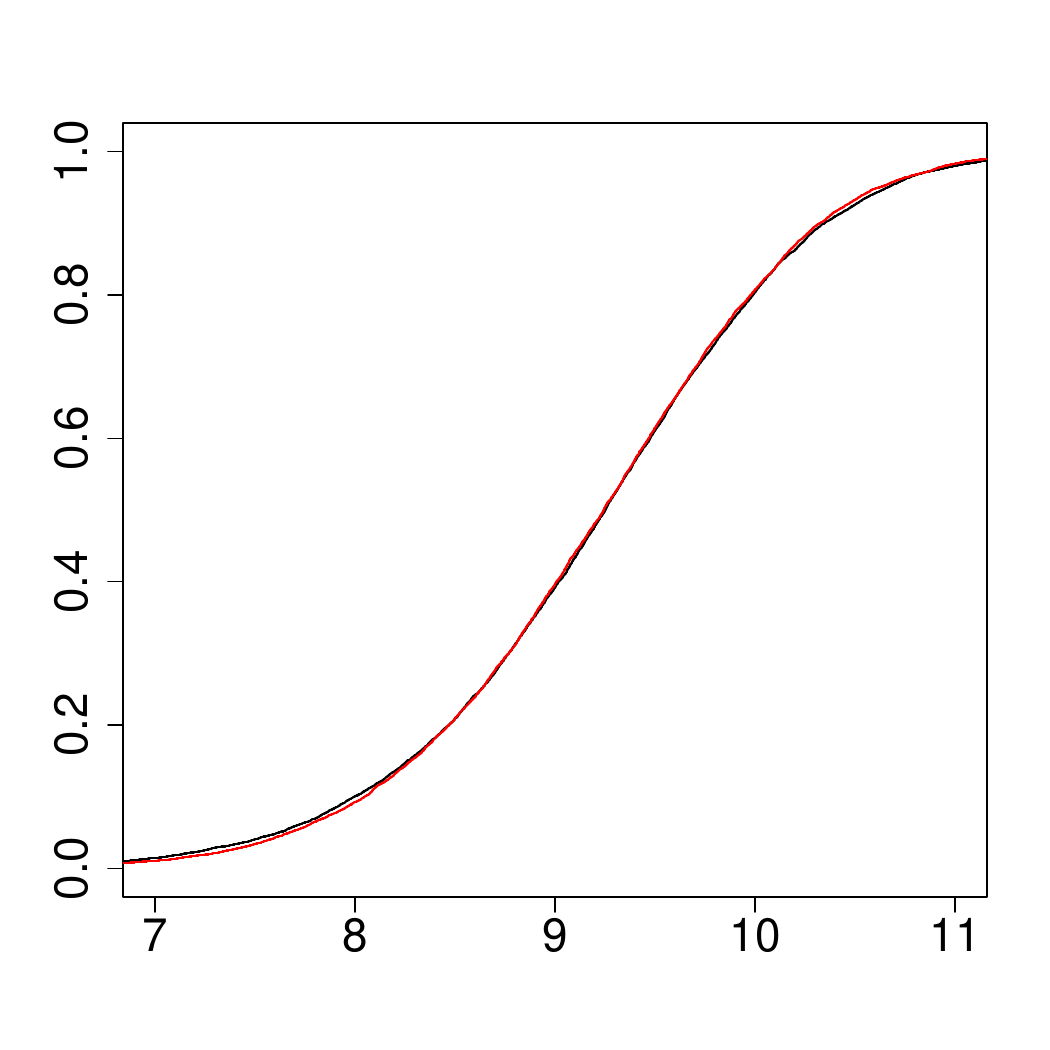}\\[-0.6cm]
        \mbox{\tiny $\ln(|Q|)$}
      \end{tabular}\\
      \mbox{\tiny Graph (c)} &
      \mbox{\tiny Graph (c)} &
      \mbox{\tiny Graph (c)} \\[-0.4cm]
      \begin{tabular}{@{}c@{}c@{}}
        \raisebox{2.0cm}{\rotatebox{90}{\tiny cdf}}
        \includegraphics[width=4.4cm]{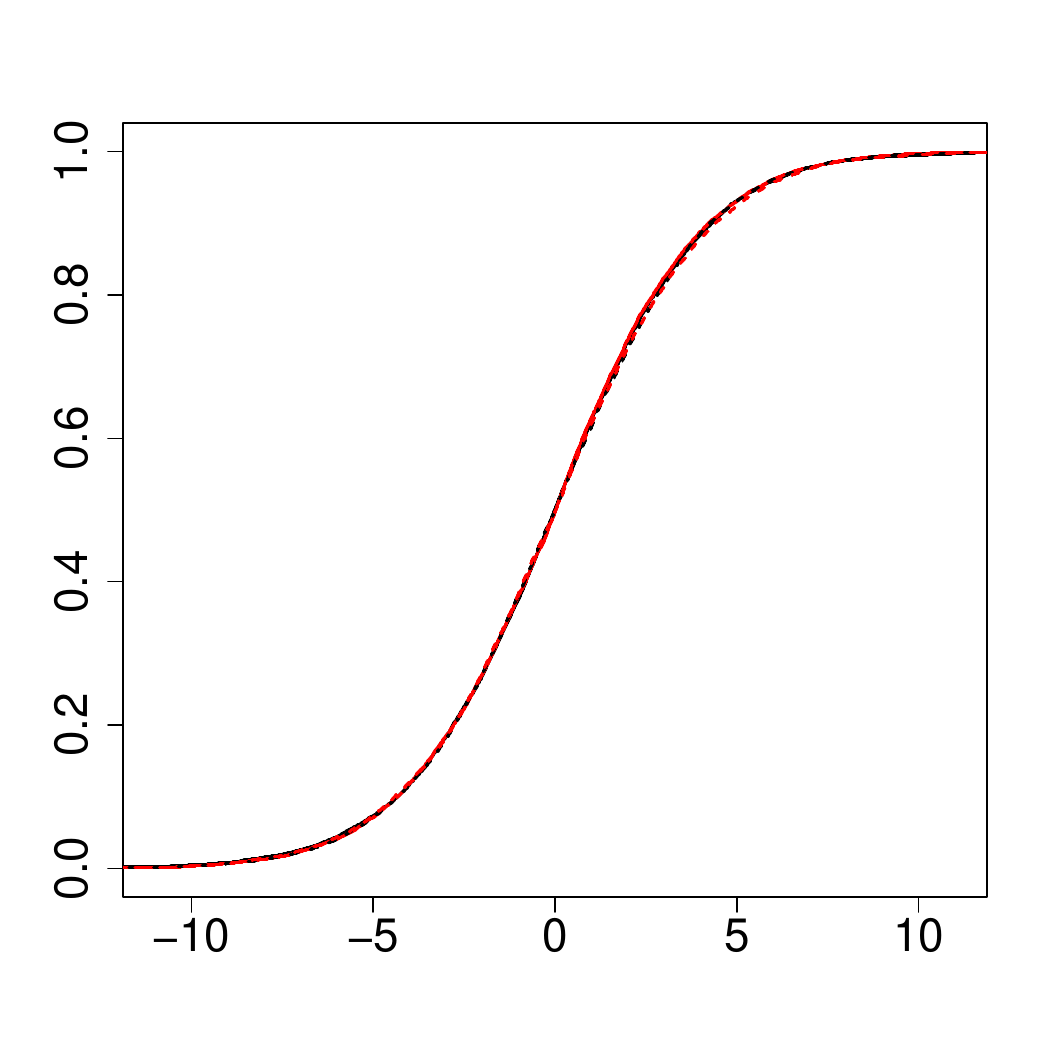}\\[-0.6cm]
        \mbox{\tiny $Q_{24}$}
      \end{tabular}&
      \begin{tabular}{@{}c@{}c@{}}
        \includegraphics[width=4.4cm]{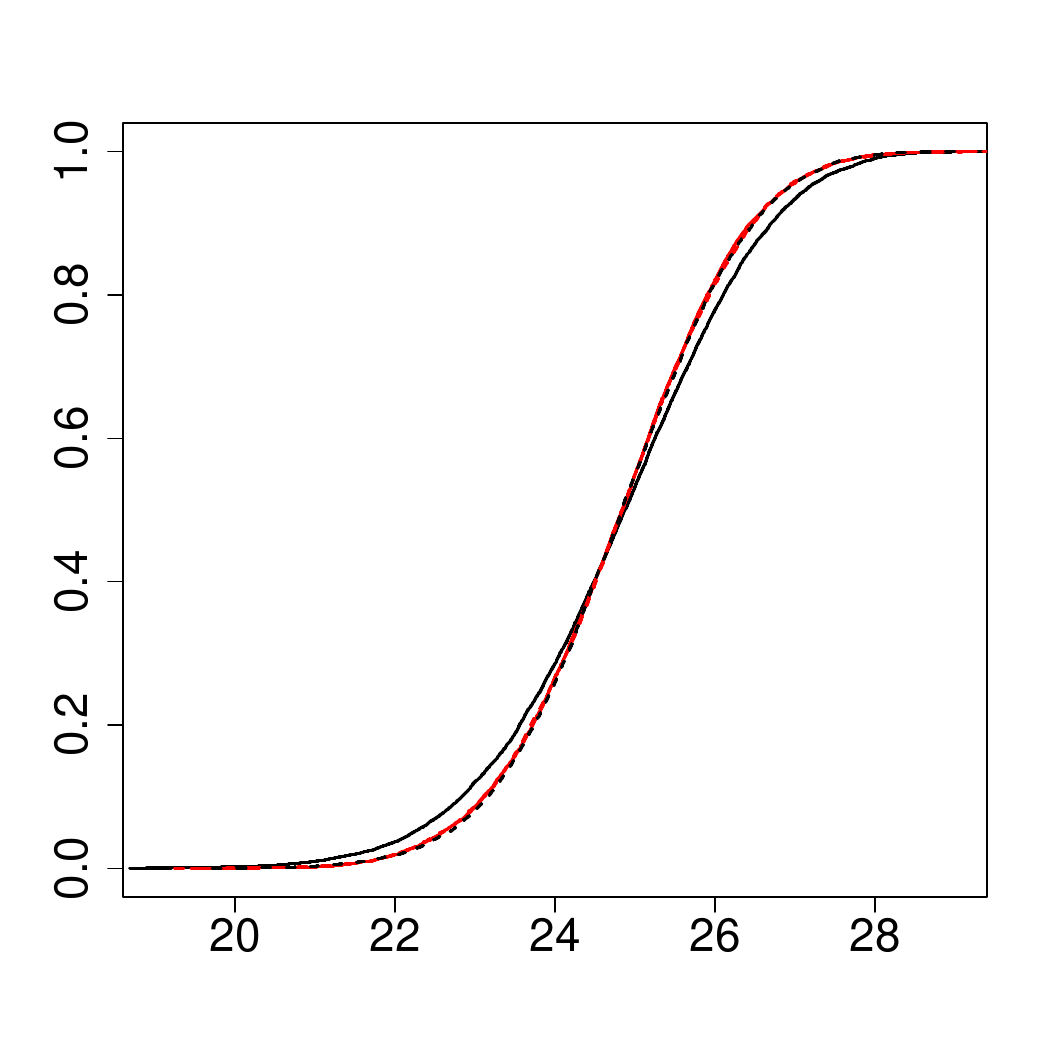}\\[-0.6cm]
        \mbox{\tiny $\ln(\mbox{tr}(Q))$}
      \end{tabular}&
      \begin{tabular}{@{}c@{}c@{}}
        \includegraphics[width=4.4cm]{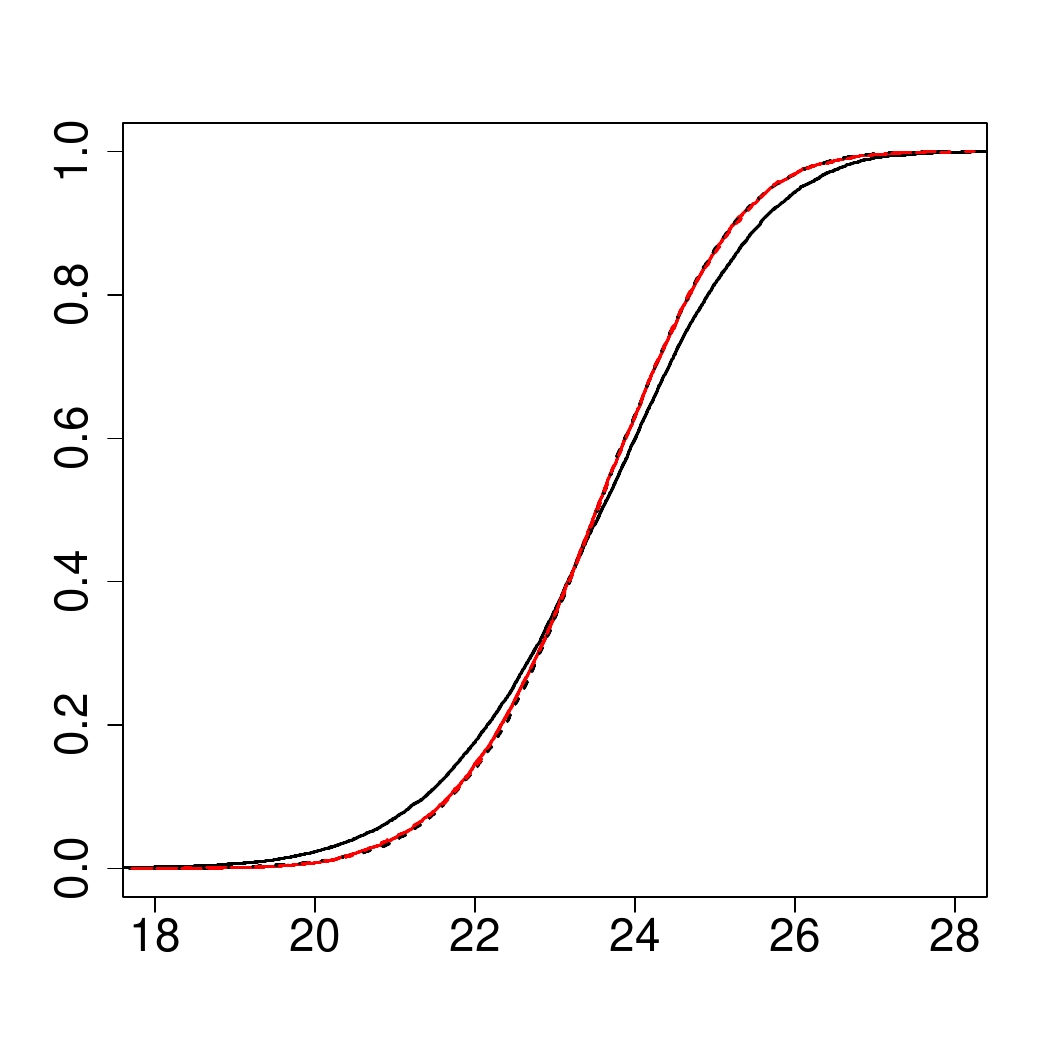}\\[-0.6cm]
        \mbox{\tiny $\ln(tr(Q))$}
      \end{tabular}\\
      \mbox{\tiny Graph (d)} &
      \mbox{\tiny Graph (d)} &
      \mbox{\tiny Graph (d)} \\[-0.4cm]
      \begin{tabular}{@{}c@{}c@{}}
        \raisebox{2.0cm}{\rotatebox{90}{\tiny cdf}}
        \includegraphics[width=4.4cm]{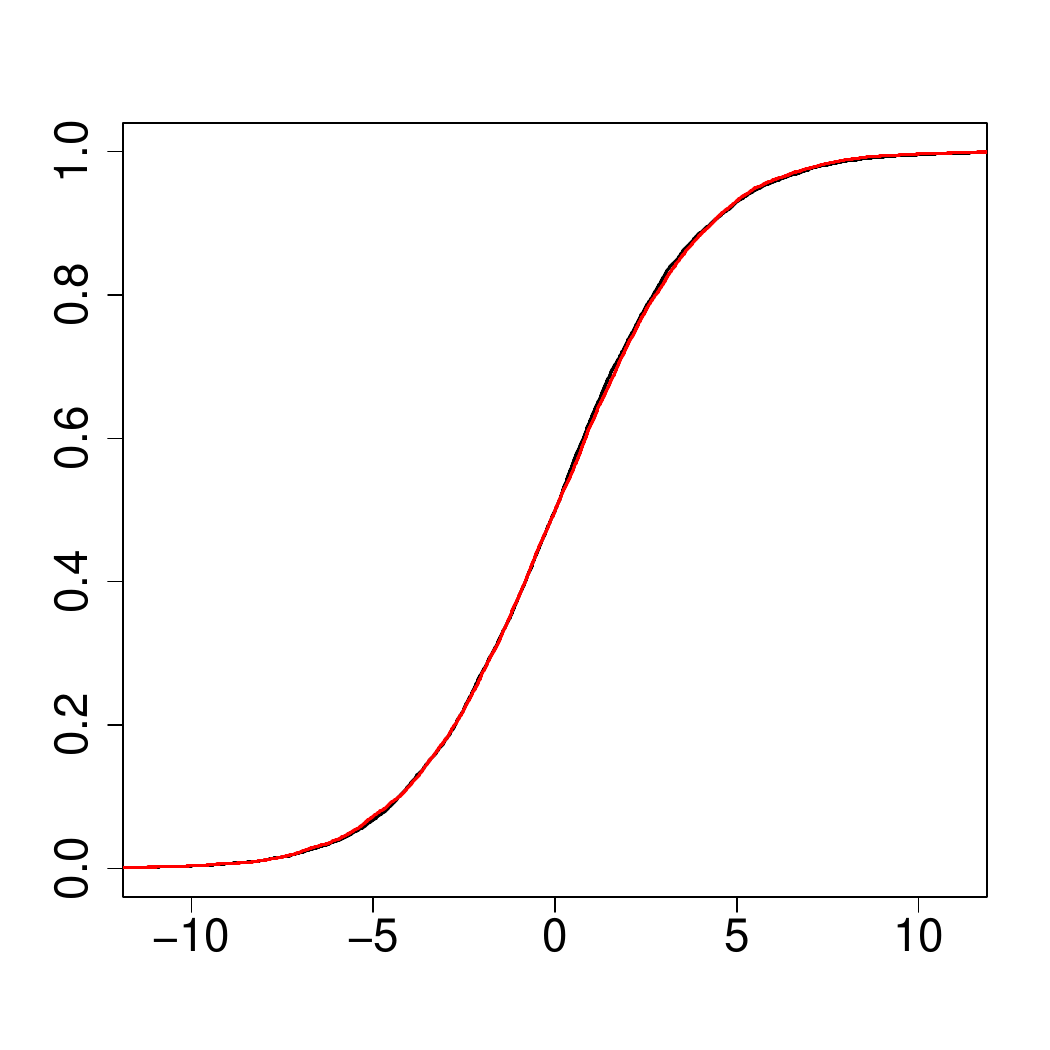}\\[-0.6cm]
        \mbox{\tiny $Q_{24}$}
      \end{tabular} &
      \begin{tabular}{@{}c@{}c@{}}
        \includegraphics[width=4.4cm]{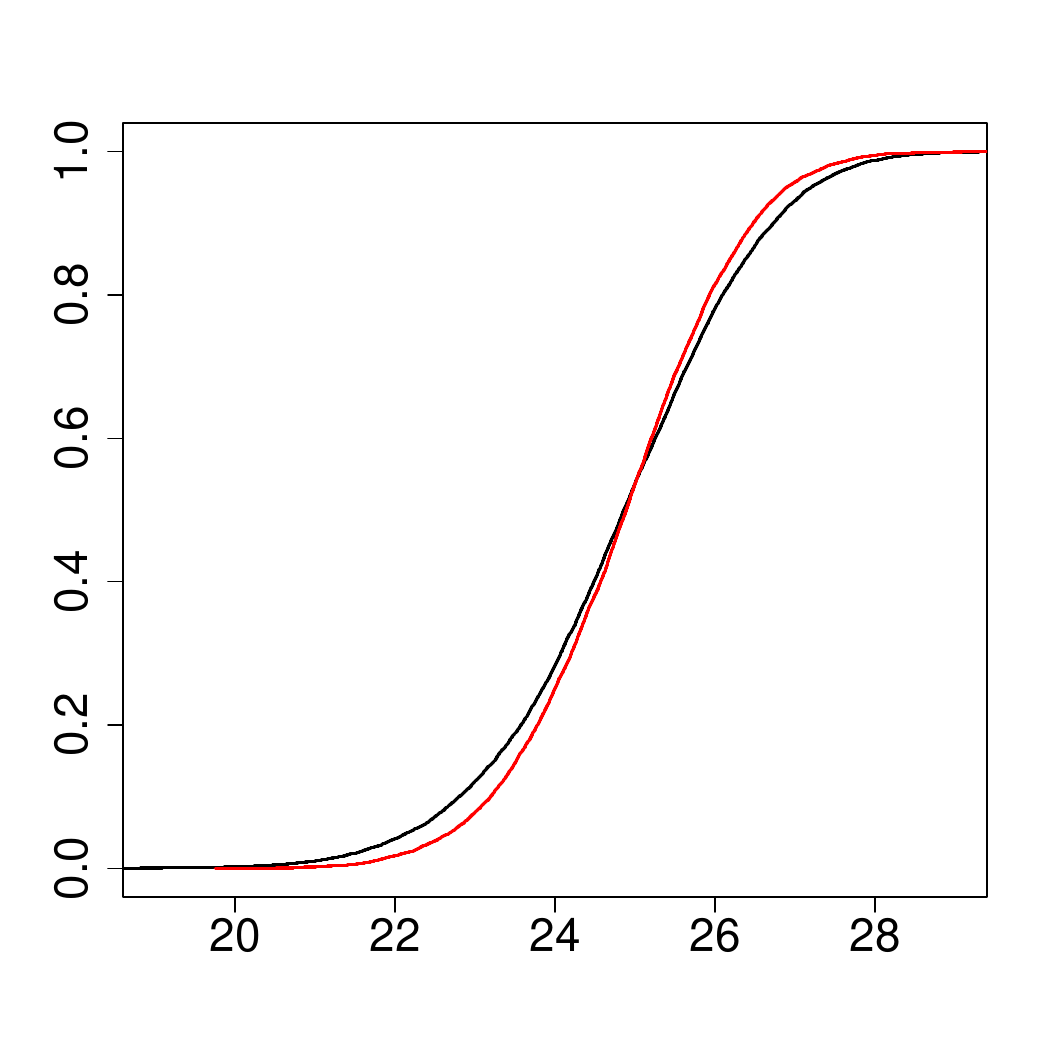}\\[-0.6cm]
        \mbox{\tiny $\ln(\mbox{tr}(Q))$}
      \end{tabular}  &
      \begin{tabular}{@{}c@{}c@{}}
        \includegraphics[width=4.4cm]{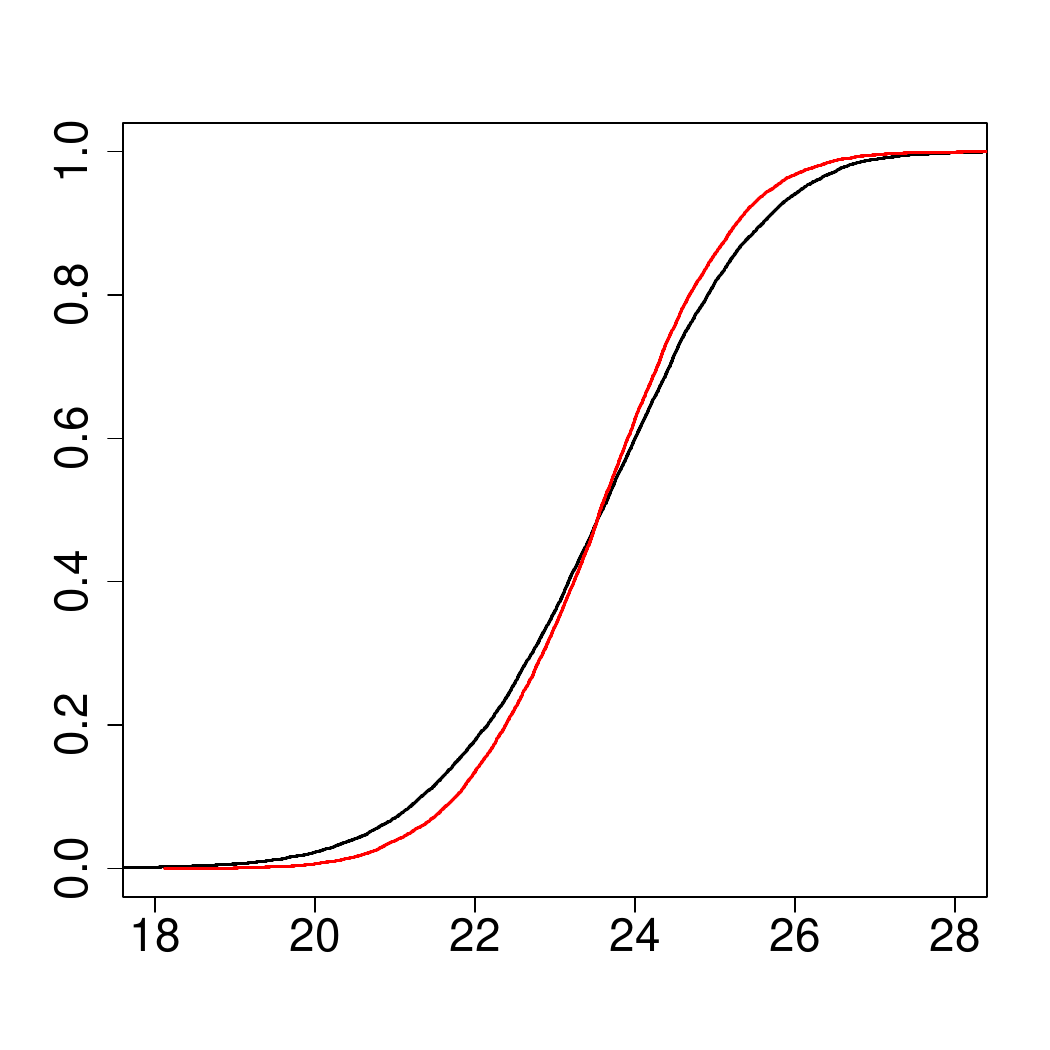}\\[-0.6cm]
        \mbox{\tiny $\ln(|Q|)$}
      \end{tabular}
    \end{tabular}
  \end{center}
  \caption{\label{fig:Cdf}Empirical cumulative distribution functions of scalar functions
    of the simulated $Q_i^{(0)}$ and $Q_i^{(r)}$ matrices. The black and red curves are based
    on the simulated $Q_i^{(0)}$ and $Q_i^{(r)}$ matrices, respectively, and solid and dashed curves
    are for the claimed sampler and for the exact sampler, respectively. In the left column
    the scalar value used is the value of element $(2,4)$ of the simulated matrix $Q$, the middle
    column is for the logarithm of the trace of $Q$, and the right column is for the logarithm
    of the determinant of $Q$. From top to bottom, the rows shows results for models based on
    the graphs in Figures \ref{fig:graph}(a), (b), (c) and (d), respectively.}
\end{figure}

For a closer study of a potential difference between the distributions of $Q_i^{(0)}$ and $Q_i^{(r)}$ for
the four models and two simulation algorithms, we compute also
the empirical cumulative distribution functions (cdfs) for the same three scalar functions of
$Q_i^{(0)}$ and $Q_i^{(r)}$ as used above. Results for the models based on each of the four graphs
in Figure \ref{fig:graph} are shown in Figure \ref{fig:Cdf}.
The plots in the left column contains the empirical cdf for
element $(2,4)$ of $Q_i^{(0)}$ in black, and the empirical cdf of the same element of $Q_i^{(r)}$ in red.
For the plots based on decomposable graphs, solid and dashed curves are again used for the results based
on the claimed sampler and the exact sampler, respectively.
Correspondingly, using the same colour and solid/dashed coding, the plots in the middle column show the empirical
cdfs of $\ln(\mbox{tr}(Q_i^{(0)}))$ and $\ln(\mbox{tr}(Q_i^{(r)}))$, and the plots in the right column
show the empirical cdfs of $\ln |Q_i^{(0)}|$ and $\ln |Q_i^{(r)}|$.

In the plots in the upper row of Figure \ref{fig:Cdf} the difference between the four estimated cdf curves
are so small that the curves are visually indistinguishable. If zooming in on specific parts of the curves very 
small deviations can be observed at some places. One should note that this corresponds well with the lack of
any observable trend in the corresponding plots in Figure \ref{fig:caseABTrace}. The plots in the second row
contain only two curves, as this is results based on a non-decomposable graph. At first sight it is again
difficult to see any differences between the curves, but when zooming in some small deviations can be observed. It is,
however, difficult to judge whether these small differences represent real differences between the theoretical
cdfs, or are just an effect of Monte Carlo error. In the plots in the third row there is again four curves,
as this is based on a decomposable model. In the middle and right plot in that row,
the solid black curve is clearly different from the other three curves.
One should note that the solid black curve corresponds to a distribution with a higher variance than for the
other three estimated cdfs, which corresponds well with what we observed in the
corresponding trace plots in Figure \ref{fig:caseCDTrace}. In the bottom row we observe a similar effect as
in the plots in the row above, which again should not come as a surprise after having seen the plots in
Figure \ref{fig:caseCDTrace}.

\subsection{Hypothesis test results}
The preliminary exploratory analysis results presented above indicate that the claimed general sampler
is not producing samples from the specified G-Wishart distribution,
whereas, as expected, there is no indications in these results that the exact sampler is not simulating correctly.
Still, however, we now present the results of the formal hypothesis test presented in Section \ref{sec:test}.
We again consider a G-Wishart distribution for each of the four graphs in Figure \ref{fig:graph}. For the two
decomposable graphs we evaluate both the claimed sampler and the exact sampler,
whereas for the two non-decomposable graphs we of course only use the claimed sampler.
We do five independent hypothesis test runs for each case.

Based on our
observations in Figure \ref{fig:Cdf} we for both the $p=4$ and $p=10$ node graph models use 
\begin{align}
  h(Q) = \ln |Q|
\end{align}
and, since the differences seem to be most notable in the tails of the distributions, 
\begin{align}
H(T) = |\mbox{Quantile}(\{t_{i1},i=1,\ldots,s\},0.1) - \mbox{Quantile}(\{t_{i2},i=1,\ldots,s\},0.1)|;
\end{align}
where $\mbox{Quantile}(A,p)$ denotes the $p$'th empirical quantile of the elements in the set $A$.

As we choose the $h(\cdot)$  and $H(\cdot)$ functions based on our observations in Section
\ref{sec:prelim}, we of course can not
use the simulated values we explored in that section in our hypothesis tests. So we simulate new $Q_i^{(0)}$
and $Q_i^{(r)}$ values for the hypotheses tests, using the same simulation setup and the same values for
$s$ and $r$ as given in Section \ref{sec:setup}. In the hypothesis test we generated $q=999\,9999$
re-sampled versions of the test statistic, as described in Section \ref{sec:test}.

Table \ref{tab:pvalues}
\begin{table}
  \caption{\label{tab:pvalues}Obtained p-values for each combination of the four graphs in Figure
    \ref{fig:graph} and the two simulation algorithms considered.}
  \begin{tabular}{l|c|c|c|c}
    & Graph (a) & Graph (b) & Graph (c) & Graph (d) \\ \hline
    Claimed sampler &
    \begin{tabular}{@{}l@{}} $0.654$\\ $0.166$\\ $0.777$\\ $0.358$\\ $0.263$ \end{tabular} &
    \begin{tabular}{@{}l@{}} $0.0161$\\ $0.00404$\\ $0.00769$\\ $0.0772$\\ $0.0394$ \end{tabular} &
    \begin{tabular}{@{}l@{}} $10^{-6}$\\ $10^{-6}$\\ $10^{-6}$\\ $10^{-6}$\\ $10^{-6}$ \end{tabular} &
    \begin{tabular}{@{}l@{}} $10^{-6}$\\ $10^{-6}$\\ $10^{-6}$\\ $10^{-6}$\\ $10^{-6}$ \end{tabular}\\
    \hline
    Exact sampler &
    \begin{tabular}{@{}l@{}} $0.344$\\ $0.578$\\ $0.802$\\ $0.479$\\ $0.731$ \end{tabular} & &
    \begin{tabular}{@{}l@{}} $0.562$\\ $0.0258$\\ $0.233$\\ $0.617$\\ $0.841$ \end{tabular} &
  \end{tabular}
\end{table}
gives the five p-values obtained in each of the six cases considered. When using the claimed sampler
for the models based on the graphs in Figure \ref{fig:graph}(c) and (d) all p-values are equal to $1/(q+1)$,
which is the smallest possible value. So these p-values give an overwhelming evidence that this
algorithm is not producing correct samples from the specified G-Wishart distributions. Also for the
model based on the small non-decomposable graph in Figure \ref{fig:graph}(b) the obtained p-values give a
very strong indication that the claimed sampler is not doing what it is intended to do. As expected,
the p-values obtained for the exact sampler do not suggest that
the generated samples are not from the specified distributions. Perhaps more surprisingly, the p-values
produced when sampling by the claimed sampler from the model based on the graph in Figure
\ref{fig:graph}(a) neither show any sign of the generated samples to be from a wrong distribution.
Of course, that we do not reject $H_0$ does not necessarily mean that $H_0$ is correct, it may just
be that we have not enough power to detect that $H_0$ is incorrect. We therefore re-run the whole procedure,
for all six cases with $s$ increased by a factor of ten, i.e. with $s=100\, 000$. We do not report all
the results of these runs here, but mentioned that then the five p-values using the claimed sampler
for the model based on the graph in Figure \ref{fig:graph}(a) were $0.000055, 0.0690, 0.000213, 0.00111$ and
$0.0143$. So the
claimed sampler is clearly not producing correct samples in this case either, it is
just that the difference between the specified distribution and the distribution from which the
simulation algorithm is sampling
is so small that we do not have enough power to detect the difference when $s=10\, 000$.
Naturally, the p-values obtained from the long
$s=100\, 000$ runs using the claimed sampler for the model based on the graph in Figure \ref{fig:graph}(b)
were even smaller than the ones reported in Table \ref{tab:pvalues}. The p-values obtained from the long runs
with the exact sampler, however, still give no indication that the algorithm is not
generating correct samples from the specified distributions.

\section{\label{sec:closing}Closing remarks}
We have formulated a Markov chain Monte Carlo hypothesis test that one can use to test whether a proposed simulation
algorithm is correctly generating samples from a claimed distribution. We have then used this
hypothesis test setup to check the claimed sampler for a G-Wishart distribution proposed in
\citet{art188}, and the conclusion is that the proposed simulation algorithm is not producing
samples from the specified distribution.

As discussed in the introduction, the small p-values obtained in the study discussed above for the
\citet{art188} algorithm can either come as a result of the algorithm not producing correct samples
from the specified G-Wishart distribution, or from a bug in our implementation of the algorithm.
To reduce the chance of a bug in our implementation, the two authors of this article independently
implemented the algorithm of \citet{art188}, one in python and one in Matlab. Both implementations
gave similar p-values, so we are confident that the error is in the algorithm itself and not in the
implementation(s). It should also be mentioned that the implementation of the proposed algorithm
does not require many lines of code.

Whenever we use a hypothesis test for observed real data, it is important to distinguish between the concepts 
of statistically significant and practically relevant. If we just have a sufficiently large sample
size we will be able to identify even a very small difference, also one which is of no practical relevance,
to be statistically significant. When we check a proposed
simulation algorithm as we do in this article, the situation is different. The claim that a proposed simulation
algorithm is correctly simulating from a specified distribution is best considered as a theorem, and a theorem is
either correct or incorrect. Even a small error in a theorem is practically relevant. However, at least for
the reasonably small graphs we have considered in this study, the differences between the specified
G-Wishart distributions and the distributions that the algorithm of \citet{art188} is generating samples from
seem to be reasonably small, so in the lack of any other algorithm simulating correctly from the G-Wishart
distribution one may argue that the \citet{art188} algorithm is an acceptable approximation. When including such
an approximate algorithm inside an MCMC setup, like the exchange algorithm, it is however hard to know
how much even a small error in the sampling from the G-Wishart distribution is producing in the resulting limiting
distribution of the Markov chain.

\section*{Funding}
  The first author acknowledges support from the Centre
  of Geophysical Forecasting (Grant no. 309960) at the
  Norwegian University of Science and Technology.


\bibliography{mybib}       

\end{document}